# Triadic Conceptual Structure of the Maximum Entropy Approach to Evolution

**Carsten Herrmann-Pillath* and Stanley N. Salthe****

* East West Center for Business Studies and Cultural Science, Frankfurt School of Finance and Management, Sonnemannstraße 9-11, 60314 Frankfurt am Main, Germany

Email c.herrmann-pillath@fs.de

** Biological Sciences, Binghamton University, Vestal, New York, USA

Email ssalthe@binghamton.edu

**Abstract**

Many problems in evolutionary theory are cast in dyadic terms, such as the polar oppositions of organism and environment. We argue that a triadic conceptual structure offers an alternative perspective under which the information generating role of evolution as a physical process can be analyzed, and propose a new diagrammatic approach. Peirce's natural philosophy was deeply influenced by his reception of both Darwin's theory and thermodynamics. Thus, we elaborate on a new synthesis which puts together his theory of signs and modern Maximum Entropy approaches to evolution. Following recent contributions to the naturalization of Peircean semiosis, we show that triadic structures involve the conjunction of three different kinds of causality, efficient, formal and final. We apply this on Ulanowicz's analysis of autocatalytic cycles as primordial patterns of life. This paves the way for a semiotic view of thermodynamics which is built on the idea that Peircean interpretants are systems of physical inference devices evolving under natural selection. In this view, the principles of Maximum Entropy, Maximum Power, and Maximum Entropy Production work together to drive the emergence of information carrying structures, which at the same time maximize information capacity as well as the gradients of energy flows, such that ultimately, contrary to Schrödinger's seminal contribution, the evolutionary process is seen to be a physical expression of the Second Law.

## 1. From dyadic to triadic conceptual structure

Evolutionary theory is haunted by a number of problems which emerged in its defining stages, and have persisted to the present time, such as the relation between ontogeny and phylogeny, the relations between genotype and phenotype, and the relation between organism and environment. In their research strategies, evolutionary theorists have tended to vacillate between polar approaches, such as, for example, gradualism versus saltationism in the broadest sense. In many respects, as in Robert Wilson's terminology (Wilson 2004: 68ff.), these oppositions can be interpreted using the 'internal richness' and 'external minimalism' format. Different approaches to evolution differ in whether they adopt an 'internal richness' position, such as in the gene-biased position, which sees genetic information as carrying the exclusive determinants of development, minimizing the role of external factors, or whether they reject one or both extremes, such as in the developmental systems approach (Oyama 2001), which continues to maintain the internal richness view, but rejects external minimalism, thus positing that biological information is contextual, that is, embodied in the larger structures of cells, organisms and even local biomes.

The conceptual trouble with dualisms is that they vacillate between privileging the epistemological positions of externalism, focusing on constraints, and internalism, focusing on generativity, depending on the aspect of evolution that is in focus. They may even fail to distinguish between the epistemological and ontological dimensions of an issue, uncritically assuming, for example, that the scientific observer is able to position herself in an external position such that possible internal positions become irrelevant, thus effectively eliminating the distinction between internal and the external. This also obscures the material fact of their own position as mediating observers by assuming a direct mapping between reality and concepts. For example, with reference to development, the NeoDarwinian view focuses internally on the locus of biological information in the ontological dimension, but at the same time is externalist with reference to the process of selection in the epistemological dimension, so that the forces of evolutionary change are seen to be located outside of the genome even though change must materially originate internally (Reid 2008). This reflects a failure to distinguish between ontological and epistemological internalism vs. externalism in the treatment of the notion of information -- that is, the question of the location of operative information and the question as to which stance of the observer the information refers. In order to keep our subsequent argument within the confines of a paper, we avoid even more fundamental issues that would arise if we were to consider that even the distinction between epistemological and ontological dimensions is itself a problem pointing toward more radical approaches, such as anchoring both dimensions internally in the observer (Matsuno and Salthe, 2002).

In the present paper, we propose that the difficulties with these questions have resulted from the fundamentally dualist oppositions in which the discussions have been framed. Ontological dualisms permeate the field, as with ‘genetic cause and phenotypic effect’, or ‘genetic sender and phenotypic receiver’. This is related to the belief that biological phenomena can be fully explained as efficient-causal, that is, as mechanistic processes (Ulanowicz 1997: Chapter 2). In the mechanistic worldview, there is no need to distinguish between different possible positions of observers in the treatment of information. This has, however, produced perennial debates about foundational issues such as the distinctions between units of selection, units of heredity and units of evolution.

Neither polar opposition is entirely satisfactory, and so we need some ‘meta’ perspective to dissolve them. Herein we propose a triadic structure to evolutionary change, in which polar oppositions are revealed to be two modes of approach to the same fundamental reality, somewhat like the wave-particle dualism in physics during 19$^{th}$ century debates. Correspondingly, biologists debate the nature of biological information, whether it is ‘particularistic,’ i.e. manifested ontologically in genes, or whether it is more holistic, i.e., ontologically manifest in complex living systems, of which genes are only one part (Godfrey-Smith and Sterelny 2008). In a triadic approach, this debate is found to be a dual perspective view. The required triadic structure can be based on the theory of signs developed by Charles Sanders Peirce (for an accessible collection of his most important works, see Peirce 1992, 1998). We will link as well his broader views on the stochastic nature of reality, and on the roles of final, as opposed to efficient causation in evolution. In referring to Peirce’s views, we mainly build on the synthesis offered by Stone (2007), who puts these views into the more explicit context of modern analytical philosophy.

In figure 1, we show the polar oppositions resolved into two modes in the fundamental triadic structure of reality envisaged by Peirce: This structure is a static snapshot of an evolutionary process in which information about an object, while not directly accessible epistemically, is generated within a system of interpretance (Salthe 2009) via sequences of interpretants informed by an evolving sign (sometimes referred to as ‘representamen’ in Peirce’s later works, a notation that has been accepted by many semioticians, using instead ‘sign vehicle’, but rejected by Stone 2007: 19, 55, whose usage we follow in the present paper, if only for reasons of consistency and expediency). However, the triadic structure of object, sign and interpretant could be interpreted in two different ways, ending up in those polar dyadic discursive oppositions.

- One way is to conflate sign and object, thus assuming a direct accessibility of objects by the reactive system of interpretants. In this 'particularistic mode' efficient causality is the framework of explanation, and this perspective also tends to adopt atomistic ontologies which gravitate towards the presumption that causal processes relate to certain fundamental entities, such as elementary particles or genes (for programmatic statements, see Wilson 1998: 53ff., 297, or von Baeyer 2003: 11ff.). In this view, the distinction between different observational standpoints becomes irrelevant, because if sign and object are conflated, it appears to be possible for observers to get direct access to 'reality,' independently of the epistemological position from which the object is approached. Interestingly, and perhaps unexpectedly, this truncation of the triadic structure has resulted in serious epistemological problems in foundational physics, where the distinction between the observer's states of knowledge and the physical object has been blurred, resulting in endless debates over the Copenhagen view on quantum mechanics, where the ontological status of 'randomness' still awaits final clarification (Jaynes 2003: 327ff.; Penrose 2006: 782ff.; Faye 2008).
- The other way is to conflate sign and interpretant, thus positing that objects are only accessible from the perspectives of particular systems of interpretance, such that there is no way to conceive of objects independent from context. This view neglects the fact that the sign is co-produced by both the system of interpretance *and* the object. This is a holistic mode, as, for example, in theories about the co-evolution of organisms and environment, which, again interestingly and unexpectedly perhaps, might also be formulated in a mechanistic fashion, because the conjunction of different causalities can only emerge in the triadic structure (see, for example, the theory of 'niche construction,' Odling-Smee, Laland and Feldman 2003). However, in this view, all potential positions of observers are equally valid, so that there is no way to arrive at a canonical unified picture of 'reality,' – a Peircean desideratum -- such as, in the aforementioned example, when trying to overcome the principled distance between the epistemic reconstruction of a niche by the external observer and the internal position of the evolving biological systems that results in niche construction (Brier 2008: 169ff., referring to Reventlow's ethological theory).

**Figure 1: The two modes in the Peircean triadic structure of semiosis**

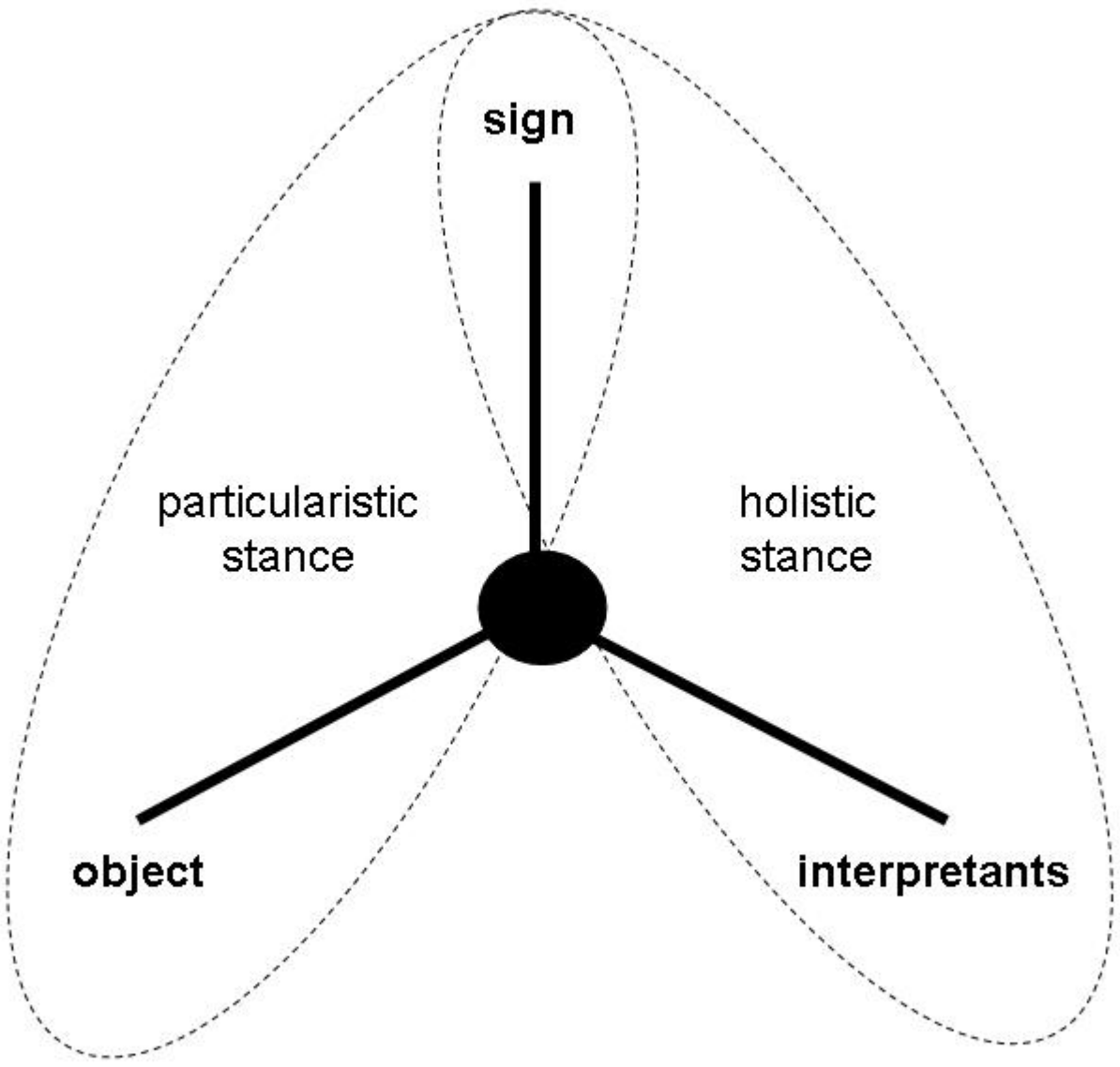

In modern science, the co-existence of the two modes was suppressed by the social construction of the ‘experiment,’ in which the relationship between observer and object was standardized in a way that conflates sign and object (Salthe, under review). However, in many sciences, such as the evolutionary sciences, the experimental construction has been difficult to implement or even impossible to achieve. Therefore, it is absolutely necessary to make the triadic structure explicit in order to clarify contentious issues in sciences confronting complex systems. The triadic structure of explanation allows for the simultaneous analysis of the two modes, and the notion of ‘sign’ obtains a central role, as it mediates two relations: one is the efficient-causal relation with the object that stimulates generation of the sign, and the other is the final-causal relation with the interpretants, which determines the larger role of the sign in the observational system in which the object is embedded.

In taking this stance, we argue that a particular methodological step is necessary. This is to adopt a naturalistic approach to semiosis – that is to say, the position according to which ‘interpretants’ are taken to be physical phenomena (upon which, for example, ‘minds’ supervene) and which can therefore be applied to all aspects of living systems. Our approach is not, however, identical to certain positions in the current ‘biosemiotics’ discourse because, for example, there can be different interpretations of the ‘mental’ with reference to the concept of the ‘meaning’ of signs, depending on whether, in the end of the day, a Cartesian dualist ontology is implied (Hoffmeyer 1999, Brier 2008), which has also been the case in panpsychistic approaches (Rensch 1974; for an overview of the alternatives, see Seager and Allan-Hermanson 2007). We follow Stone's (2007: 156ff., 301ff.) interpretation of Peirce in taking 'mental' phenomena to emerge from the semiotic process, and therefore they cannot serve as an explanation of that process. Yet, our view implies that 'mental' characteristics could emerge in all physical structures that manifest semiosis, without implying Cartesian dualism. Recently, several attempts have been presented offering naturalistic versions of Peircean semiosis, which differ in details, but less in substance (Vehkavaara 2002, El-Hani et al. 2006, Southgate and Robinson 2010, Herrmann-Pillath 2010, Salthe, under review). We build on these approaches, but add a substantial extension, which relates particularly to Peirce's interest in the interplay between randomness and regularity. This is to bring thermodynamics and statistical mechanics into the picture.

The established way of naturalizing semiosis is to understand the concept of the interpretant to be a 'response' or a 'function' in a semiotic system, basically similar to the teleosemantic approach to meaning as function (Millikan 1989; Macdonald and Papineau 2006). This entails reference to a function of functions, so to speak, hence to a hierarchy of functions, which is central to realizing that signs themselves cannot be seen as having functions (Stone 2007: 159ff., 172). We argue that the (subsumptive) hierarchy (Salthe 2002):

{PHYSICAL TENDENCY {CHEMICAL AFFORDANCES {BIOLOGICAL FUNCTIONS }}}

represents the fundamental properties of living systems, which can be viewed as exploiting information to harness and dissipate energy (the physical tendency -- for convergent views, see, for instance, Lahav et al. 2001, Elitzur 2005). Both information processing and energy dissipation can be effectively viewed from the standpoint of thermodynamics and statistical mechanics. A major aim of this paper is to establish a direct conceptual bridge between the latter -- hence the concepts of information and entropy -- and Peircean semiosis, in order to shed new light on the aforementioned external / internal polar oppositions in evolutionary theory. We expect that this conceptual synthesis may offer an opportunity for resolution of other open questions in evolutionary theory, which, however, we will only allude to in the final section of this paper. Our aim here is to establish fundamentals.

We start with an analytical summary of one recently published attempt at a naturalistic semiotics, by Robinson and Southgate (2010). We expand on this by means of an analysis of the underlying triadic structure, proposing clarification by way of a new diagrammatic exposition. Dovetailing with R&S's analysis of the origin of life, we generalize the argument by offering a semiotic analysis of the universal model of autocatalytic cycles as found in, e.g., chemistry, ecology and social systems. This allows for linking the Peircean semiotic analysis with thermodynamics. Then we continue by placing the structure of statistical mechanics into the triadic framework, so that we can apply basic conceptual principles from statistical mechanics to semiosis. We show that in a naturalized semiotics, semiosis encompasses both the information theoretic Maximum Entropy Principle and the physical Maximum Entropy Production Principle.

## 2. Naturalizing Peircean semiosis

### *2.1 Basics of triadic conceptual structures: Efficient and final causality in evolution*

A conceptual framework for a naturalistic semiotics has recently been proposed by Robinson and Southgate (2010) (R&S) who build on Short's (2007) account of Peirce's theory of signs. The basic idea is that 'interpretants' in Peirce's approach are linked to a 'general response', *R*, in any system. This corresponds to use of the term 'function' in teleosemantics, which implies that responses are embedded in larger systemic contexts, such as biological functions in an organism. A response, *R*, to an object, *O*, is informed by a sign, *X*, that represents *O* to the responding system. R&S do not use the term 'information' here because, according to their understanding, it suggests only the Shannon information concept. However, we think this conceptual barrier is unnecessary, as the natural notion of information always implies semantic information -- that is, data with meaning in context (Floridi 2003, 2007). Furthermore, Shannon information is merely a quantitative measure of information. For us the information concept will be useful in recognizing that the dualism of sign and object plays precisely the role of generating information about the object for an observing system. In other words, in our naturalized view, the semiotic relation between object, sign and response is an inference mechanism: The sign mediates a response that, if that response meets certain criteria of proper functioning, it carries information about the object embodied in the corresponding physical interactions (as in the 'inference devices' in the sense of Wolpert, 2001, 2008). This information-generating role has been thematized by Peirce in his distinction between 'immediate object' and 'dynamical object', and the idea that the deployment of sequential interpretants is a process of approximating the 'reality' of objects, relative to the contexts in which the interpretants are embedded, as well as in regard to the interest, qua functions, of the observing system. This is why Peirce's theory of signs and his theory of inference form a unity (Atkin 2006), which we could also project into our analysis of evolution.

This statement needs careful consideration: We cannot say that the sign carries information about the object unless this information relates to an interpretant (Stone 2007: 172). For example, even in the standard case of smoke and fire, smoke is a direct (efficient) causal effect of fire, and only carries information if there is at least a potential observer that makes the inference from smoke to fire. However, this implies that the information is embodied in the observing system's global response, and not in the sign exclusively. In Peircean semiotic analysis there is no 'view from anywhere' which is implied in the particularistic stance of standard science discourse. For example, if the observer is interested in whether there is fire because she would flee in order to survive, then the response, running away, embodies the information that there is fire, and not the smoke as such. This is also evident from the fact that the same physical entity, such as smoke, can be a different sign, dependent on the interpretant (for some, smoke indicates fire, for others, smoke indicates human company, depending on context and interest, see Short 2007: 189).

The basic triadic structure of semiosis as naturalized is depicted in figure 2, following the notation introduced by R&S, modifying and extending it on the basis of a related diagram of theirs. This differs in several important respects from other triadic diagrams that can be found in the literature (e.g. El-Hani et al. 2006, Brier 2008, Salthe, 2009). Actually, this is a diagram of two triads, one of which goes via the *Q* at the center and a more encompassing one passing through *X*. In the first, *Q* represents the physical mechanisms that connect object, sign and response. R&S define *Q* as a property of an entity -- which we will just call 'living system' here -- of undergoing a change of state when causally impacted by *O*, such that the response *R* occurs. For example, *Q* contains all the physical and neuronal chains of causes and effects that would connect smoke with the movements of a person fleeing from the fire. All interactions of *Q* can be separately described mechanistically, whence our use of the term 'naturalization.' *Q* is a blackbox that contains several different mechanisms, which, however, become integrated into the semiotic process via the interpretants (at R). Insofar as the identification of the boundaries of that blackbox cannot be determined physically (for example, there may be many other causal effects of smoke that are irrelevant for the action of running, yet are also physically connected in an indistinguishable way), *Q* cannot be isolated from the embedding semiotic relations, because the scope of relevant causal effects can only be defined relative to the dynamic relations between all of *O*, *X* and *R*.

**Figure 2: A naturalistic view on the Peircean semiotic triad**

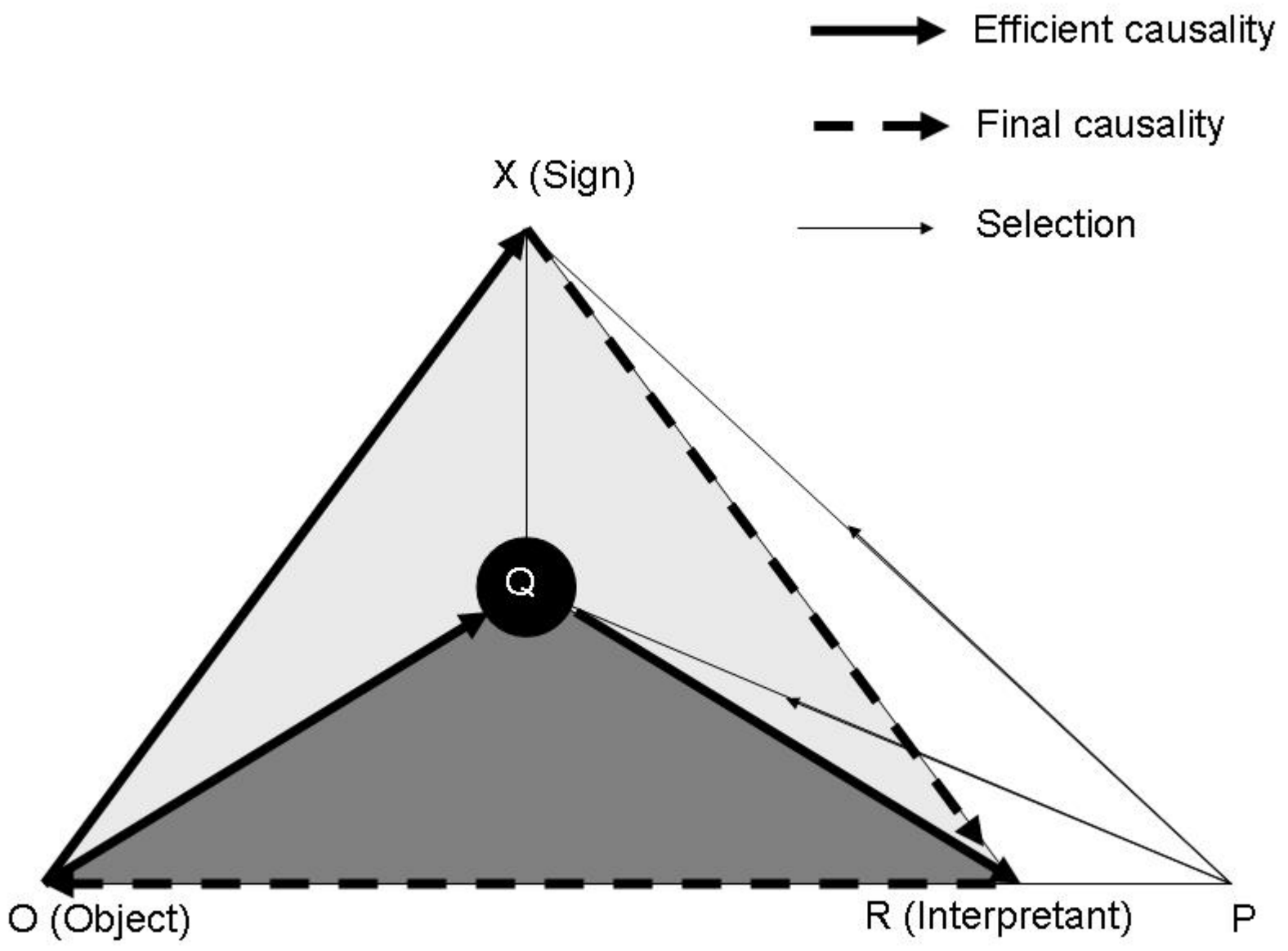

So, the blackbox *Q* contains the mechanisms by which an object causes some effect(s) on a system, which are twofold, corresponding to the two triads. One is the effect that is mediated via the sign, such as, e.g., neurophysiological mechanisms of perception, involving *R*. These are the mechanisms by which the physical phenomena emanating from smoke are identified perceptually as 'smoke', as being separate from other physical phenomena in the environment. The other is the effect that the object has on the larger scale organism, and which relates to the larger, more general, function of the response, such as running (here, away from fire). It is important to realize that *Q* has been affected by natural selection, which is indicated by the thin arrow that relates *Q* to *P*. *P*, following R&S, represents the pull of the response by its larger function ('purpose') in the organism. R&S define natural selection as a process in which particular functionings are selected for a 'general type of outcome,' such as 'avoiding dangerous fire,' which allows for a multitude of different possible activities. The central insight depicted in fig. 2 lies in the observation that the evolution of both *X* and *Q* are guided by natural selection, and that the physical content of *Q* is entangled with the evolution of *X*. This is the central mechanism by which natural selection results in generating information about *O*.

The bottom or inner triangle, in dark shading, is embedded within the larger one, partly in dashed lines. We propose to understand this embeddedness as referring to 'supervenience.' Thus, *X* supervenes upon *Q* in the same way as mental processes supervene upon neuronal processes (McLaughlin and Bennett 2008). Supervenience has the important property of multiple realizability: The same sign, in its relation to the response, could be released by many different states or aspects of the object. As we shall see, this is the ultimate reason why signs generate information within the system of interpretance as it confronts an object. In preparation of things to come, we notice that this relation of multiple realizability of *X* corresponds to the distinction between microstates and a macrostate in statistical mechanics, such that a large number of microstates of an object correspond to one macrostate causally connected with the object.

The difference between the two triangles, *O-Q-R* and *O-X-R*, in figure 2 lies in that the larger one describes a process guided by final causality, as in Short's (2007: 136ff.) precise definition, though being materialized via efficient causality, too (that is, for instance, the sign *X* connects with *R* efficient-causally, if this relation is taken in isolation). Diagrammatically, therefore the broken arrow X-R supersedes a solid arrow (not shown for reasons of clarity in the diagram). This is because the relation between sign, response and object is a generic one, which directly connects to mechanisms producing general types of outcome. That is, viewed from the inner triangle, the sign is a particular in Peirce's sense, hence a token, but viewed from the outer, partly dashed triangle it is a Peircean 'general,' hence a type. This differentiation of two levels, one supervening on the other, will presumably have resulted from natural selection, which can be viewed as a process that maps relations between object and sign into a 'general' in Peirce's terms -- that is, a general type of outcome according to Short. It is the sign that enables a semiotic system having the property *Q* to manifest a response *R* by which the system is enabled to relate with the object *O* under the perspective of the 'purpose' *P*. So, in the supervening level of the semeiotic relation as solid arrow (the object causing the sign) is related with two broken arrows, which, via the *R*, lead back to the object, thus reversing the direction of the causal forces, with the object becoming causally influenced by the interpretant. This is meant to display final causality.

Therefore, in the naturalization of semiosis, the notion of 'purpose' is pivotal, and comes close to the notion of 'proper function' in teleosemantics. In Short's approach, continued in R&S, a purpose is associated with selection, which is implicitly viewed as selection *of* something *for* some purpose, with the latter designating a general type of outcome. We notice that epistemologically, this general type of outcome is not a simple observable, but can only be reconstructed conceptually by means of a theory of the larger system in which *P* is embedded, such as, for example, differential reproductive success in the NeoDarwinian theory of evolution. In other words, in Peircean terms, selection maps particulars into generals. Reproduction can subsume innumerable variations of particular traits. This general type of outcome is a criterion of selection. For another example, high speed movement might be a general type of outcome in a predator-prey system, based on any number of anatomical or behavioral adjustments within the prey. The point is that this outcome can be achieved in many different ways, so that the result of selection can be described as final causality that *supervenes* on various underlying processes, each animated by efficient causality. This means that, in natural selection there is a sequence of causal events that leads towards the differential reproduction of organisms with traits such as higher speed, which sequence could be described in terms of efficient causality. But the direction taken by the process is determined by the property of higher relative speed, which is a final cause, subsumed by the yet more general final cause of reproduction.

To summarize, in our interpretation, there are the following core components of the R&S approach, building on Short's more fundamental conceptual distinctions.

- *R* relates to an object such that it contributes to the generation of a generalized type of outcome, thus *R* mediates a purpose. This means that the object must have a causal effect on *R* that would be relevant in terms of the purpose *P*. This effect works via *Q*.
- *X* also relates to *R* in a mechanistic way This relation is a mapping from the efficient causal relation between *O* and *X*, which works via the co-evolution of *X* and *Q* under natural selection, resulting in the supervenience of a formal-causal relation *X*-*R* to the mechanistical relation *X*-*R* .
- The different relations mentioned previously are reflected in a property, *Q*, of the system that manifests in the response *R*. This property has been selected in the light of purposes *P*.
- The embeddedness of the two triangles, both building on relations of efficient causality, manifests the supervenience of final causality on efficient causality, such that the purpose *P*, via the intermediation of the evolving *X*, guides the generation of information about the object *O* in the evolutionary process.

**Figure 3: The system of interpretance**

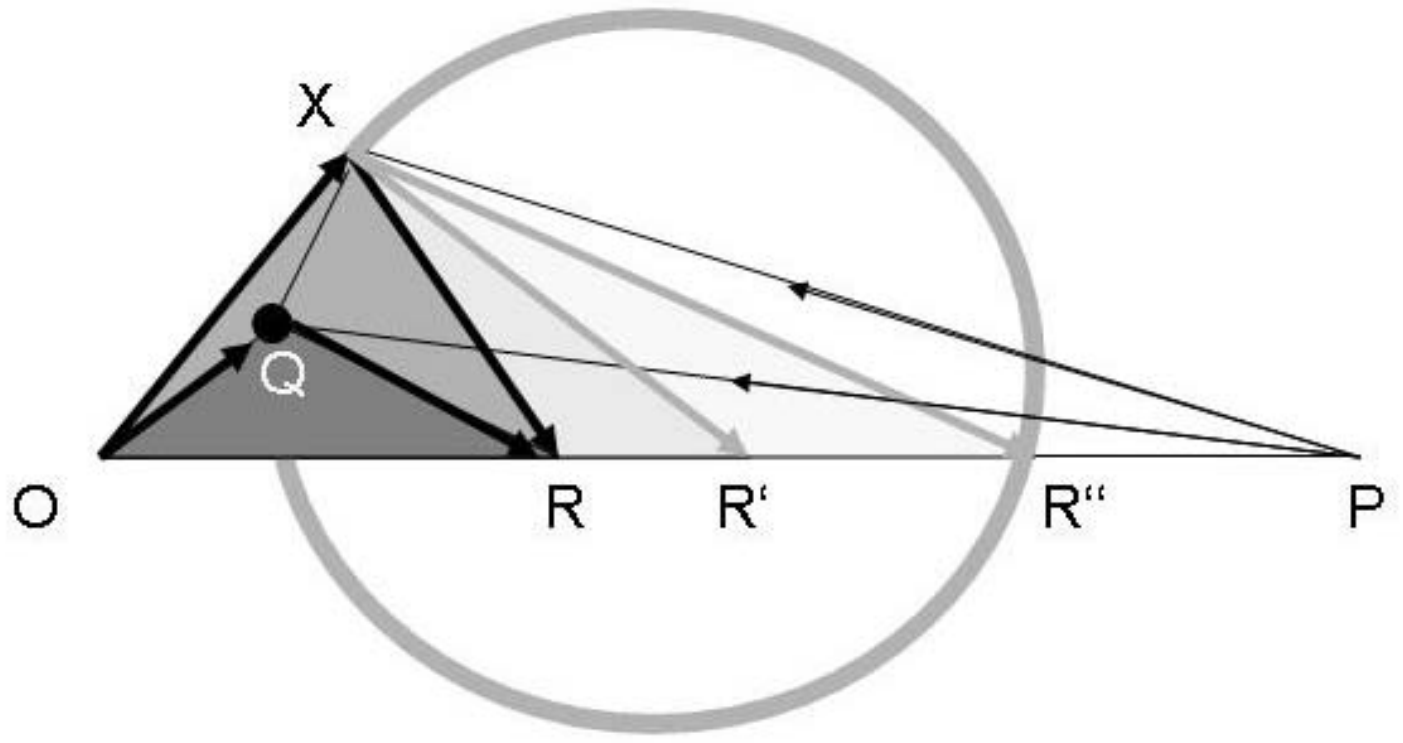

This conceptual scheme is extended in figure 3, showing further additions to figure 2, and also simplifying the relations of different causalities, in order to highlight the result from adding the notion of hierarchically ordered purposes. Every purpose is a nested structure with different degrees of generality, as in:

{P {R {R' {R'' { etc. }}}}

So, it is not sufficient to refer only to the most general purpose, *P*, such as 'differential reproductive success.' R&S use the term 'proximal purposes' in order to make this point, while emphasizing that *P* and *R* should not be conflated, because otherwise there would be no possibility of poor, or non-functional, responses and functionings of signs, which is central for the validity of this approach, as well as for Peircean 'fallibility', required also in teleosemantics. Yet, we can say that proximal purposes connect to a chain of hierarchically ordered responses. For example, the purpose of an engine can be to transmit power, which is general. The more limited purpose might be to move a car, or it can be viewed as to move the wheels of a car. Using the subsumptive hierarchy we get:

{TO POWER {A CAR {BY WAY OF ITS WHEELS }}}

Clearly, the different purposes are also increasingly specific functions. If we transfer this observation onto the semiotic analysis, then we have responses / interpretants which can give rise to other responses / interpretants. For example, an animal may respond to the shape of another animal with flight. Flight is related to the generalized purpose of survival. Yet, the immediate causal loop may just relate the observation of the shape with a movement of the limbs as a response. This response alone is not sufficient for flight. For example, there must be another response that is directing the movement away from the perceived threat. Thus, a complex reaction such as a movement can be dissected into hierarchies of functions, which can be viewed as increasingly 'more proximate' purposes. So, movement of limbs has the purpose to move in a certain direction, and the movement toward a certain direction serves the purpose to avoid a perceived predator.

Now, we posit that such hierarchies of purposes make up a 'system of interpretance' (SI, Salthe, 2009). Systematicity implies two things here. Firstly, as depicted in fig. 3, there is a chain of responses qua interpretants of signs which are reflected in certain mechanisms that in complex systems will have evolved as the effects of selection. Secondly, the system has boundaries with respect to other systems. With the concept of system of interpretance, we are supplementing the R&S model. For example, an SI can be an organism, which manifests many different interrelated functions related to various purposes. A purpose can be determined in different ways, depending on the hierarchical level considered. Thus, an organism is part of an ecosystem which constitutes a higher level SI in a compositional hierarchy (Salthe, 2002). In this higher level (here larger scale) system, we can identify higher-level purposes which embed the organismic functions. Similar lower-level responses can also have different higher-level purposes, which in turn relate with even broader SI's, such as, in the limit, might be envisaged in 'Gaia' approaches. This increasing scale (in the compositional hierarchy) or generality (in the subsumptive hierarchy) of context also reflects the fact that living systems ultimately are related with a limited number of more universal purposes, such as differential reproduction of a population, the generation of energy flows in an ecosystem, etc. Thus, the SI provides connections between a systemic response and multiple purposes, in this way extending the R&S approach.

### *2.2 Semiotic analysis of autocatalytic cycles*

In order to prepare the ground for the argument in the next section, we propose that in a naturalized semiotics, the archetypical case of systematicity is the autocatalytic cycle, which we analyze in semiotic terms. Again, this follows the R&S approach (and see also Herrmann-Pillath 2010) in using the origin of life scenario as a reference case. However, we claim that this reference case can be viewed as just one example of the more general model of autocatalytic cycles (or hypercycles, in particular), which applies in many different contexts and across systemic levels, reaching from chemistry to human social systems (Maynard Smith and Szathmary 1995, Padgett 1997, Padgett et al. 2003, Odum 2008). Hence, it provides an analytical backbone for generalizing the semiotic triad. We notice, but refrain from elaborating further on this point here, that this argument can even be further generalized onto dissipative structures as physical phenomena, which corresponds to recent attempts at building foundational physics on an information-theoretic basis (see e.g. Zeilinger 1999, Lloyd 2006, Salthe under review).

In Ulanowicz's (1997: 41ff.) outline of the autocatalytic cycle, the following features are essential. First, there is a process that generates an effect on another process which increases the rate of activity of this second process. Then, the second process also generates a similar effect on another process, which may result in a linked sequence of processes. Finally, there is a process in the chain of them that contributes to increasing the rate of activity of an earlier process in the chain. Ulanowicz shows that the resulting overall process cannot be exclusively understood in terms of efficient causality, even though every step in the process can be viewed as efficiently caused. This is because the internal interdependence introduces a distinction between external and internal selection on the overall unit of interrelated processes. Internal selection happens because any variation in a single processes that enhances the overall level of activity will be leveraged by the interaction, and in the reverse case, reductions will be also leveraged, such that an internal directedness will emerge that Ulanowicz calls 'growth enhancing', with 'growth' here referring to growth in the overall throughput of the cycle.

If one considers the case of a population of autocatalytic cycles, this implies that those which increase their overall rate of activity will increase their share of resources in a population of them. This results in external selection of units of cycles. Therefore, internal directedness translates into external directedness on the population level, a relation which Ulanowicz calls 'asymmetric.' This expression directly corresponds to Stone's (2007: 115) notion of 'anisotropic processes' as opposed to mechanical processes. From this follows what Ulanowicz calls the 'centripetality' of the autocatalytic process. Centripetality means that the cycles with higher rates of activity attract more resources and energy from their environment to feed their growth. In other words, autocatalytic cycles tend to maximize power throughput in the sense of Lotka's (1922a, b) principle of natural selection. Centripetality implies competitive pressure on other cycles in the neighborhood, with the neighborhood being defined by the scope of the resource constraints, not spatially (Matsuno and Swenson 1999).

Autocatalytic cycles are a simple case of the emergence of a higher level in a compositional hierarchy, understood as either larger in scale (or, alternatively, greater in complexity in a subsumptive hierarchy). This is because with an autocatalytic cycle, its integrity is independent from variations of its constituents, as long as those constituents retain their integrated catalytic function with respect to other processes. This means that any element in the cycle could be substituted by a functionally equivalent (even if in unrelated ways phenotypically different) element. Therefore, cycles can be functionally identical on the higher level, even though consisting of different elements entirely.

In the analysis of autocatalytic cycles, Ulanowicz distinguishes three causal processes and compositional levels, which directly correspond to our triadic framework -- that is, can be reconstructed in terms of the embedded semiotic triads. One is the lower level of the component efficient-causal processes -- the chemical kinetics. The other is the 'focal level,' which is formal-causal in the sense that the structure of the cycle is ontologically autonomous from its individual constituent processes. We can also say that formal causation is a static view of an autocatalytic cycle. Finally, the autocatalytic cycle establishes a functional unit which underlies its growth dynamics and its developmental directionality of increasing throughput. It is, at a larger scale, a developmental trajectory (Salthe 1993). This brings in final causality, which can here be described as the tendency to maximize energy throughput, linking the cycle's activity to the universal tendency to dissipate energy gradients at the fastest possible rate as the more encompassing final cause (Schneider and Kay 1994, Niven 2009, 2010).

**Figure 4: Triadic structure of the autocatalytic cycle**

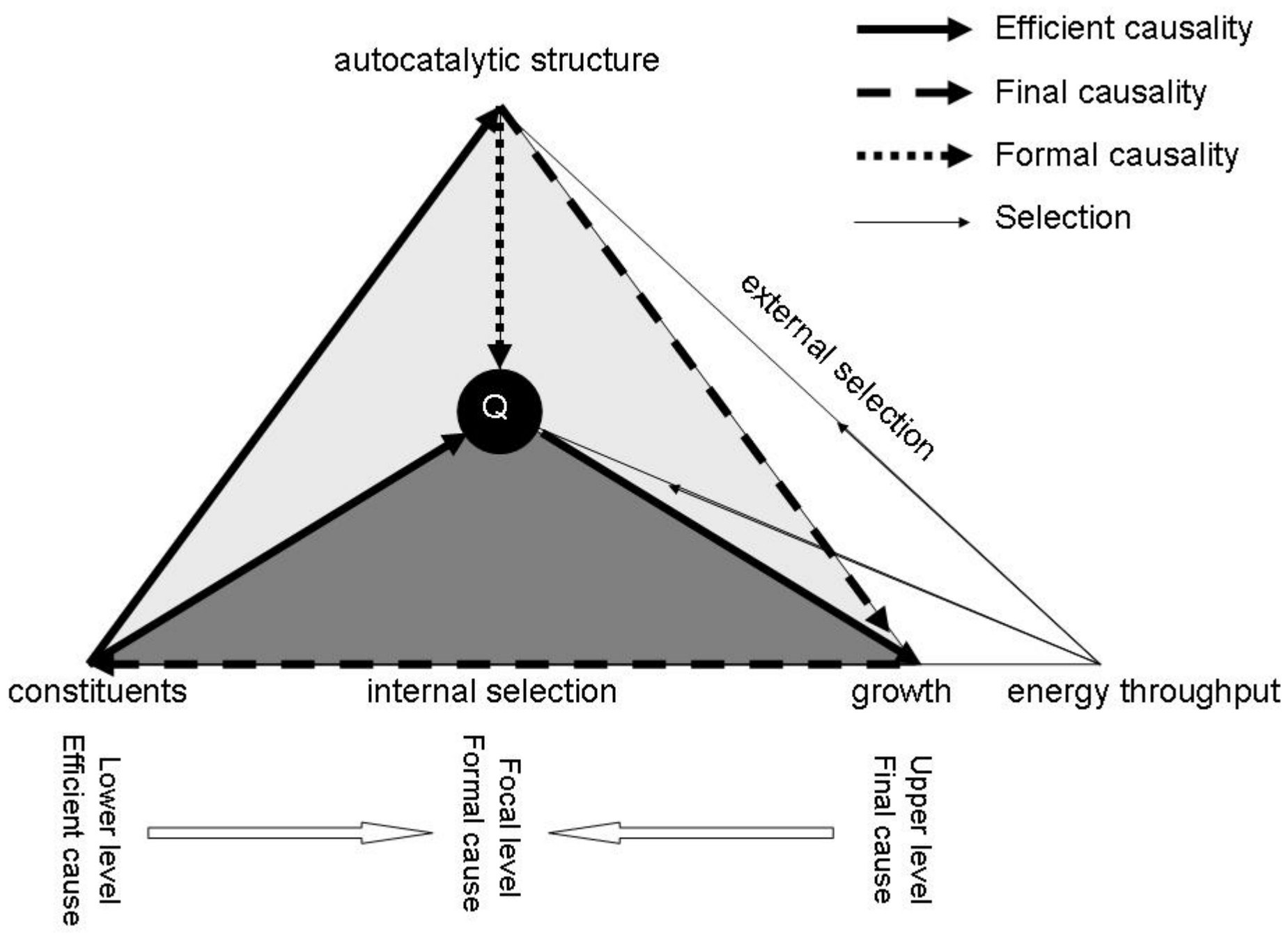

So, we end up with the following triadic diagram of the autocatalytic cycle (figure 4), which we supplement by Ulanowicz's (1997: 52) diagram of the relation between the three levels and the causal types (in the bottom). This differs from the previous diagram in explicitly identifying formal causality (represented by the vertical dotted line), working from the structure to *Q*. *Q* is now to be seen as the sequential pattern of the chemical interactions in terms of kinetics, whereas the structure (identified at its apex) is the abstract form of chemical functions, such as would described in the overall formula for the aggregate catalytic functions. This abstract pattern underlies the internal selection of components, whereas the external selection of different versions of the cycles in a population determines their relative shares of environment resources. The correspondence between the semiotic concepts and Ulanowicz's analysis of the autocatalytic cycle springs to the eye.

In which sense, then, can we understand autocatalytic chemical cycles as being semiosic? In the first place, this results from the distinction between the active parts of an involved molecule and the entire molecule, which is, in living systems, particularly pronounced in enzymatic reactions, in which the shape of a molecule obtains the central role in determining its functions. In a solution with different chemical agents, the kinetics would be determined by the molecules, but the enzymatic activity is determined only or mostly by the active sites. Thus, the active sites stand in an object-sign relation with the molecule in its entirety, insofar as the active site relates with another molecule that takes part in the catalyzed reaction. In other words, the catalyzed reaction and its product make up the response *R*, or the interpretant taken from the chemical solution in which the reaction happens, and in which aspects of molecular shape assume the role of the sign *X*. Beyond this, these relations are determined by the function of catalysis. That is, whether a molecule is a catalyst or not depends in part upon its environment. Only in the case of autocatalysis is there a direct functional interdependence between the concentration of a molecule in a solution and its catalytic effects on other linked reactions, and hence, on the concentration of its products. The joint product of this interdependence constitutes an (immediate) finality, which is essential for semiosis. We note as well that if the environment of the cycle were depleted of all but one input into the cycle, then the active site of that member of the cycle would signify for the cycle as a whole the chemical activity that it mediates, serving as a sign of it.

From another perspective, and directly referring to fig. 3, the chemical structure of the autocatalytic cycle is a property that supervenes on the underlying field of efficient causality in the sense that the cyclic interdependence only relies on the emergent and partial spatial properties of the different molecules involved (which corresponds to the general emergent relation between shape / structure and atoms in molecules, Del Re 1998, Ramsey 2000, Vemulapalli 2006). This imposes constraints in terms of final causality on the further evolution of the chemical composition of the environment in which the reactions take place, which is the relation between *X* and *Q* in semiosis. The response *R* results in growth of the rate of cycling, thus operating as an internal selection on the composition of the environment. Arguing this way, it is also immediately evident that the relation between the cycle as *X* and the components as objects, *O*, is also a relation between microstates and macrostate in the statistical mechanics sense. In the autocatalytic cycle, this holds for the response *R*, too. The autocatalytic structure is a macrostate that, in principle, allows for a multitude of different microstates on which it supervenes, and which are selected by their contribution to the response *R*, which enhances the power of the structure. In other words, the relation between macrostate and microstates allows for neutral variations, the scope of which would be determined by internal selection, both on the level of the transition from the molecular composition to the dynamical shape, and the level of the spatial matches between sites and substances (for a pertinent argument on proteins in general, see Fontana 2007).

Now, given these correspondences between autocatalytic structure, object-sign relation and micro / macro-state distinctions, the question arises if, in a comprehensive view of chemical and biological evolution, we could elaborate on the role of energy dissipation in the autocatalytic cycle as a physical correspondence to final causality in semiosis. This opens up the possibility to relate semiosis with fundamental aspects of thermodynamics.

## 3. Statistical mechanics and semiosis

### *3.1 The role of final causality in statistical mechanics*

The relevance of the Peircean approach for evolutionary theory can be derived from a fundamental insight into the statistical nature of the theory of evolution (Fisher 1958), which it shares with thermodynamics. In the words of Short (2007: 117ff.), both processes are anisotropic, hence manifest the property of directedness, which cannot be explained on purely mechanistic terms. In both cases, this directedness emerges from stochastic processes. In the case of thermodynamics, this is the expression of the Second Law. In the case of evolution, this is the result of selection, when it is viewed in a certain way. The statistical nature of selection and the closeness of evolution to thermodynamics had already been emphasized by Ronald Fisher, as well as George Price (1975), who emphasized the relation between information theory and evolution in his foundational treatment of selection (Frank 1995).

In simple terms, the commonality results from the fact that in a statistical explanation, the difference between the initial and the final state of a process under consideration does not rest on properties of the individual entities that make up the relevant population, but on certain properties of the ensemble of individuals, such as their distribution over a certain space partitioned in a particular way, or the temperature of the ensemble, which is a property of the population, and not of the individual members. So, if we state that the process will reach the most probable state, attainable by way of a maximum number of trajectories, this is independent of any force that impacts the individual particles. Similarly, if we state that evolution tends towards the realization of certain states (maximizing population fitness), these states themselves are not determined by particular properties of the individuals, but by properties of the entire ensemble of individuals and environments, such that, again, the final state could be realized by way of many different actualisations -- a maximized number of them.

Viewed in this way, the simple, yet essential conclusion follows that evolution can be seen as 'survival of the likeliest', in the sense of a multiple realizability of end states of evolutionary processes (e.g. many possible ways to fulfil a certain adaptive function, thus a general type of outcome in the sense of R&S and Short). So we see that the common characterization of evolution ending up in 'improbable states' is in fact a fundamental misunderstanding of the underlying statistical processes, because it confuses our perception of the complexity of individual living systems with the statistical properties of the populations, which is the focus of the statistical theory of selection. In fact, the distribution of biological forms represents the results of selective sorting which necessarily would be those variants of pathways of change which are the most probable ones, given the constraints, and relative to a certain context (Whitfield 2007, Dewar and Porté 2008, Dewar 2010). Simply, then, the prevalence of a certain type of organism under certain environmental conditions just shows that this particular pattern of biological organization was the most likely given the constraints (organismic and ecological) under which the selection of patterns operates. Seeming exceptions, like bizarre deep sea fishes, would from this perspective result when the selective intensity is low compared to the variability in the population. So, both natural selection and statistical mechanics build on the common foundations of statistical processes in ensembles of individuals, and hence belong to the general class of anisotropic processes.

For the statistical framework, the distinction between macroscopic and microscopic states in a state space is central. This concurs with different and distinct notions of causality, which have been in confrontation since Boltzmann conceived the foundations of statistical mechanics, and which had left him in a 'tragical impassée' (Cohen 1996). In this framework, one posits a regularity in the changes of the macrostate of a system, and argues that this process follows a statistical principle which determines that those macrostates will materialize that correspond to the most probable next microstates in a given set of possible ones. This principle is the maximum entropy principle. It is important to recognize the simple, yet often overlooked fact that in this argument there is always only one macrostate unequivocally and deterministically related with any given microstate. This differs in principle from those approaches in quantum theory that regard the macrostate itself as a probabilistic phenomenon as depicted in the Schrödinger equations. Thus, probabilistic reasoning only refers to properties of ensembles, of which single physical microstates might be a part, and which therefore cannot be seen as a physical, or more exactly, a mechanistic aspect of reality (which Bayesians such as Jaynes, 2003: 74, 411, therefore criticize as a 'mind projection fallacy').

This explanatory structure is triadic, not dyadic. In Figure 5 the triad consists of object microstates, macrostate representations, and systemic responses by way of interpretants. Recognizing its triadic structure, we immediately see that two different kinds of explanation are involved which proved difficult to reconcile in the early decades of the development of thermodynamics, when the notions of mechanistic reversibility and thermodynamic irreversibility clashed. One is the mechanistic explanation, the other is the anisotropic explanation. In fig. 5, we can analyze this in more detail, involving the different causalities that we have already related to each other in understanding the autocatalytic cycle (which, in our current argument, would itself count as an example of an anisotropic process).

**Figure 5: The triadic structure of thermodynamics**

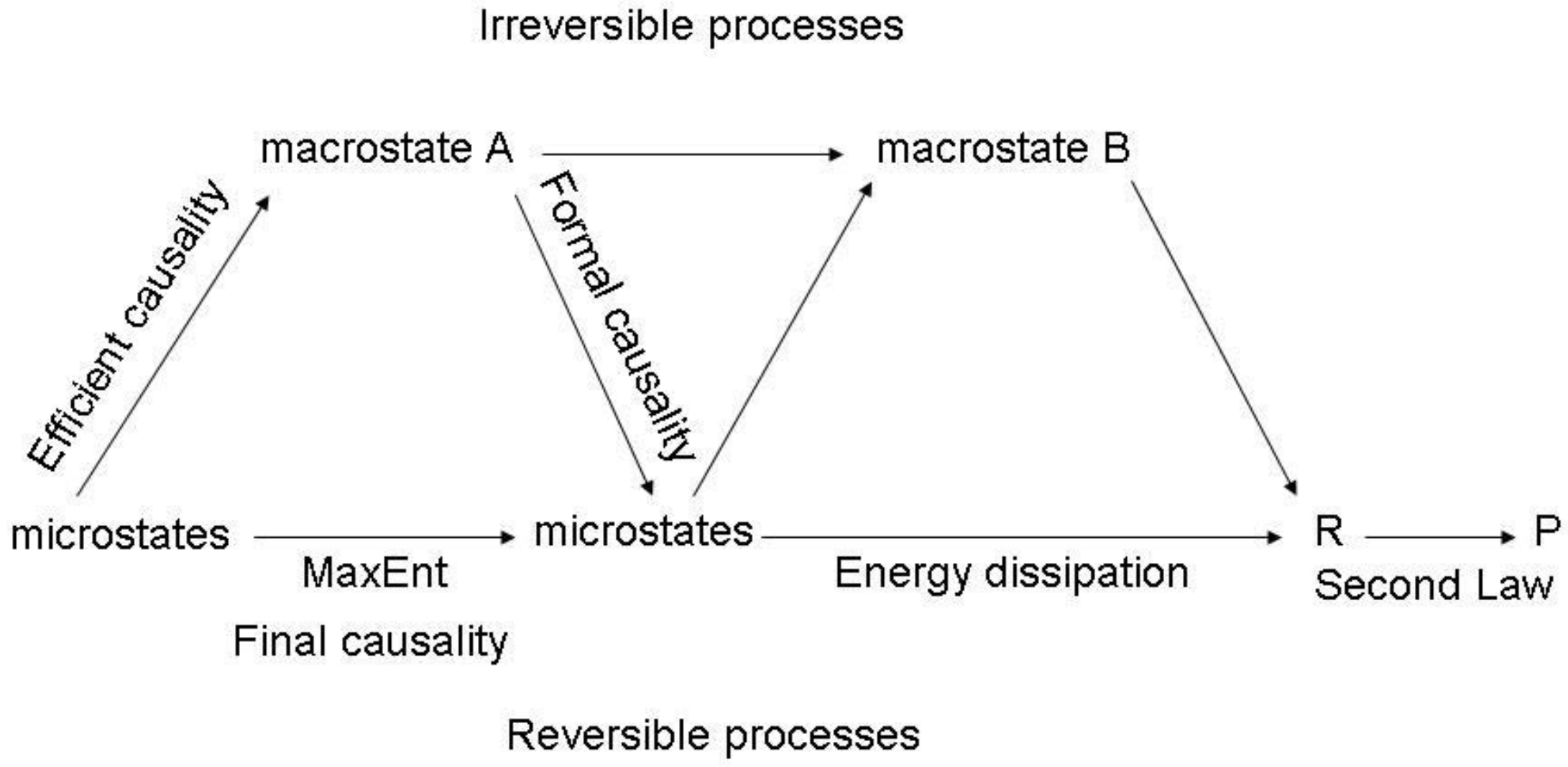

As we have already stated, every microstate corresponds to one macrostate, such that the macrostate is efficiently caused by the microstate. For example, the temperature of a gas is one <- many, as determined by the kinetic energy of the molecules. However, the transition from microstate to macrostate involves an important conceptual transition, because macrostates manifest the property of multiple realizability, hence they supervene on the microstates: Every macrostate can be realized by a number of different microstates (compare Sklar 2009). Therefore, the transition from microstate to macrostate involves a transition to a different form of causal explanation: In the relation between macro- and microstate, a probability can be assigned to any particular microstate, which is a property of the ensemble of possible microstates, without implicating that this ensemble is a physically real thing. Therefore, we can no longer think in terms of efficient causality here. The assignment of probabilities corresponds to formal causality in our analysis of the autocatalytic cycle: Thus, in an autocatalytic cycle, its dynamic structure would determine a range of possible individual actualizations of the cycling. We can now generalize, in the sense that the statistical argument puts macrostates and microstates into a formal-causal relation to each other.

Then, once the probabilistic framework is established, it allows for the formulation of a hypothesis about change of macrostate. This hypothesis is the maximum entropy principle, which states that those macrostates will be tend to be realized which could be realized by the maximum number of possible microstates. This is a purely statistical hypothesis, which does not relate to any notions of efficient causality, and hence is totally independent from the underlying physical properties of the process in question. The MaxEnt principle is universal, as long as certain fundamental formal properties apply. So the distinction between microstates and macrostates and the corresponding establishment of the formal-causal framework generates the conditions for the application of a final-causal principle, the MaxEnt principle. This is final-causal because it is anisotropic, that is, it introduces the notion of directedness in the analysis of the observed process.

The central point is that the determination of a resulting microstate cannot be based on the efficient-causal framework alone, because the directedness of change can only be determined via reference to a macrostate, which in turn refers to the probabilistic context, which is not physical. We note that the commonly used image of particles moving by efficient causality away from collisions toward their most dispersed state actually depends upon formal relations imposed by boundary conditions. Hence the transition from a purely dyadic framework (cause and effect in mechanistic terms) to a triadic framework (the additional distinction between microlevel and macrolevel) is necessary to explain the observations.

Some resulting microstates might correspond to a new macrostate, *B*, which can be predicted by application of the MaxEnt Principle on the part of the observer, who reasons that macrostate *B* will be that state in which the dispersion within the microstate has been increased. This introduces many conceptual debates about the notion of probability (for an overview, see Hájek 2010), especially with reference to the distinction between so-called ‘subjective’ and ‘objective’ interpretations. Here we meet another dyadic conceptual structure, where the debates can be resolved in a triadic framework which recognizes different kinds of causality. On the one hand, Jaynes' (2003: 290ff., 343ff.) 'Bayesian' definition of the MaxEnt principle is fully justified because the formal-causal framework is not the same type (Salthe, 1986) as an efficient-causal physical reality, and thus is set up relative to the expectations based on the knowledge of an observer. But -- and in this sense contrary to Jaynes -- this observer is a part of physical reality, such that the status of the MaxEnt Principle is not just a conceptual one, but refers to a property of the physical reality that is being observed, namely the directedness of the developmental process linking up microstates, that multiply realize respective macrostates. This directedness is 'objective,' and would correspond to a notion of propensity (which is vehemently criticized by Jaynes 2003: 60ff.). Yet, as a physical phenomenon of directedness it cannot be explained in terms of efficient causality, but only via the conjunction of formal and final causality, the reconstruction of which necessarily involves reference to an abstract observer (for instance, in the definition of the state space or the constraints, see below). In this sense, the term 'statistical mechanics' is systematically misleading, as the central explanatory scheme is not mechanistic (see Ulanowicz 1997: 24ff.).

In the debates over thermodynamics raging at the turn from the last century, the failure to distinguish between dyadic and triadic conceptual structures also found expression in the opposition between the reversibility of the micro-level effects and the irreversibility of the macro-level processes. This is basically the distinction between statistical mechanics and thermodynamics. In figure 5, this difference is visualized in the distinction between two levels of processes, the micro-level processes, which are reversible, and the macro-level processes, which are irreversible. The two levels imply two different interpretations of entropy, one in the statistical sense, and one in the sense of phenomenological thermodynamics, and hence, in the context of the Second Law. The Second Law, as has been demonstrated by Jaynes (1965), follows from the MaxEnt Principle, and hence can also be seen as a purely statistical principle (Dewar 2005). However, at the same time it refers to relations between physical magnitudes which are not statistical as such, like energy, temperature or pressure. The central physical phenomenon is the dissipation of energy, which is the transformation of energy that can exert physical work into energy that cannot. This energetic perspective introduces as well a cosmological dimension into the analysis, which is necessary to explain the relation between specific physical processes and the underlying universal trend of energy dissipation (Layzer 1988, Chaisson 2001, 2005, Penrose 2006: 686ff.).

In the triadic formulation, we distinguish between maximum entropy and the Second Law because we need to distinguish between the statistical properties and the physical properties of a process and the resulting states. The maximum entropy principle identifies the most probable macrostate, given a reference frame, which implicitly refers to an observer. The Second Law and the corresponding process of energy dissipation refers to dynamical physical activities that approach the dispersed microscopic state, either in the sense of physical process leading towards the corresponding macroscopic state, or in the sense of the microscopic flows in the direction of that state. This distinction is necessary to distinguish between the two main interpretations of the statistical mechanics foundation of thermodynamics, namely the Maximum Entropy Principle and the Maximum Entropy Production Principle. The conflation of the two triangles in fig 5 into one is not desirable, because that would seem to merge the two principles into one. Currently, there is an empirical and theoretically informed discussion about whether the former principle necessarily implies the second principle, such that both are just two sides of the same coin (Kleidon and Lorenz 2005b, Kleidon et al. 2010). This is the question that we will pursue now.

### *3.2. A semiotic interpretation of thermodynamics*

We can now relate our triadic analysis of thermodynamics to the Peircean notion of semiosis, which allows us to analyze the triadic structure more deeply. This is possible because in the Peircean approach, the notion of interpretant is a physical one, that is, it does not presuppose mental phenomena (which are seen as emergent properties of the semiotic process). The general idea will be that directedness results from a semiotic process that underlies the interaction between microstates and macrostates, and which corresponds to an inference process that is pulled by a general purpose, namely the maximum dissipation of energy. We concentrate our argument on living systems, even though the physical framework is more encompassing (Salthe in press). This focus renders the depiction of the intermediating role of natural selection more straightforward and easier to integrate into established positions in the literature.

The relationship between object, sign and interpretant we take to be a relation between a realized microstate (in thermodynamic disequilibrium), an unrealized, yet evolving macrostate (thermodynamic equilibrium) and the fluctuating microstates of a system. The current value of the macrostate is causally related to the microstates, but there is multiple realizability -- that is, the same macrostate could result from many different microstates. When introducing semiotic analysis, it is important to notice that macrostates can differ in nature, especially with regard to the location of a system's boundaries. There are macrostates which are connected to properties of a system, and there are macrostates which are caused by a system but are not properties of that system. Taking smoke and fire: smoke is a direct causal effect of fire, but may not be regarded as a property of the burning material as such, because it depends upon the entire systemic boundary, wherein smoke can be seen as a part of an entire process involving the dissipation of energy. These differences relate to the two-faced nature of a sign, which relates on the one hand with the object and on the other with the interpretants deployed by the semiotic system in reaction to it. Whether and how a macrostate relates to a microstate depends as well upon the system's deployment of its interpretants.

This view corresponds to the Jaynes view on thermodynamics in the sense that the sign has the function to generate inferences about the object via the intermediation of the system's interpretants. In the case of thermodynamics, the inferences aim at predictions of future states of an object. This inferential process is not explicit in the standard treatments of statistical mechanics concerning the probabilistic nature of the changes in the physical properties of the object. However, it becomes explicit in the Bayesian approach favoured by Jaynes, where the MaxEnt Principle is explicitly conceived as an inference process. This oscillation between two views reflects the dyadic structure of these arguments, which is made obsolete by taking a triadic view.

Comparing the triadic structures in figures 2 and 5, we can propose a further interpretation. Here we treat the relationship between macrostate *A* and macrostate *B* as a relation between a sign and its alteration as a result of system responses. The relation between macrostate and microstate is a formal-causal one in the sense of section 2.2., because, as we have seen in the discussion of fig. 5, the assignment of probabilities to the microstate does not relate to any efficient-causal property of theirs. Thus, we can interpret the macrostate as a 'sign' in the Peircean sense. In this case the differentiation of macrostates is relative to an observer that only implicitly inheres in the fundamental triad, in the Peircean sense of a 'possible' observation ('interpretability' according to Short 2007: 188). In other words, the distinction between macrostates is relative to an observer, which, however, remains hidden in the fundamental assumptions about the state space that underlies the assignment of probabilities. Here, macrostate *B* is viewed as an interpretation, or response of the system to macrostate *A*, a description which conflates observer and system in the previous sense (see fig. 1). In the Gibbs-Jaynes approach this corresponds to the experimental setting in which certain dimensions of the state space are chosen to describe the macroscopic consequences of microstate evolution. If this act of defining the experimental setting is removed from the picture by abstraction, we end up with the conflation between observer and system. In this case, thermodynamic macrostate *B* is being viewed as the interpretant in a semiotic relation to the sign, macrostate *A*.

So, we can project figure 2 onto figure 5, resulting into figure 6. This figure shows that the concept of the 'purpose' here corresponds to the MaxEnt Principle. In the Gibbs/Jaynes view, and directly corresponding to the semiotic analysis, the relation between purpose and response, (macrostate *B*) is one of inference. This means that the transition from macrostate *A* to macrostate *B* generates information about the microstate, here viewed as the semiotic object. Here, this information relates to the constraints of the system (*C*), which correspond to *Q* in the extended form of semiosis shown in figure 2. So macrostate *B* reflects the information that inheres the object, hence the constraints that operate on the maximum entropy negotiations. For example, if there is a physical space structured by a number of compartments, the equidistribution of particles in this space will directly reflect information about the structure, while the positions of the particles will attain the maximum entropy state, relative to that structure.

**Figure 6: A semiotic view on thermodynamics**

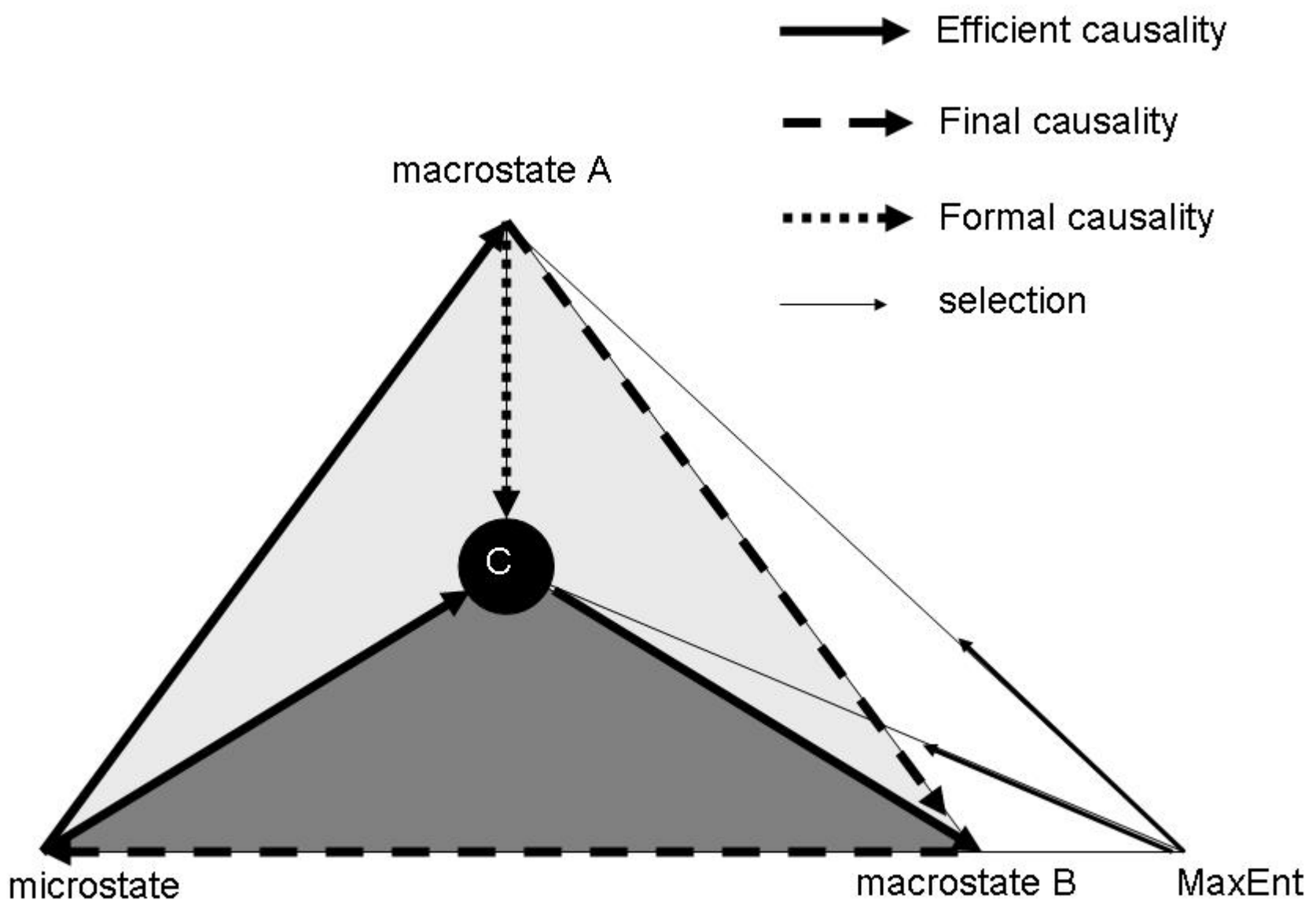

Macrostate *B* is closer to thermodynamic equilibrium than macrostate *A*. The idea that the transition from *A* to *B* generates information is also implicit in the fact that only at equilibrium can the central thermodynamic magnitudes on the macrolevel be defined. That means that the equilibrium state here corresponds to Peirce's notion of the 'final interpretant,' as compared to the previously traversed immediate and dynamic interpretants, with the latter now conceivable as a sequence of non-equilibrium states relative to the final equilibrium condition. Compared to fig. 5, fig. 6 reaches the following insights. First, the semiotic object is conceived as a physical system that shows transitions between states that are subject to constraints. These constraints are being manifested in the particular courses taken by the efficient-causal processes that lead towards macrostate *B*, which relates with the maximally dispersed microstate, while multiple realizability holds, because there is a larger number of possible other microstates corresponding to that particular macrostate. Second, the distinction between macrostates *A* and *B* is a semiotic relation that allows for extraction of the information contained in the microstate of a system. Third, this extraction works via the prediction of macrostate *B* given the macrostate *A*, which is therefore a sign that relates formal-causally with the constraints *C*. Fourth, macrostate *B* is therefore an interpretant in the sense that the information contained in the constraints *C* is extracted if and only if macrostate *B* reflects the maximum entropy state of the physical system under constraints.

In this analysis we reach two fundamental conclusions. One is that 'thermodynamic equilibrium' is not conceived as one actual physical state, which is why we do not show a corresponding microstate in fig. 6. It would only be an actual physical state if we made the observer to which the construction of the macrostates refers explicit. Equilibrium is an interpretation based on the actual physical macrostate *A*. This is the fundamental reason why the MaxEnt approach can be applied on non-equilibrium systems. Indeed, one could say that the MaxEnt approach applies to non-equilibrium systems in general. It regards the notion of equilibrium as a property that is imposed by the inference process, but has no necessary particular physical interpretation in terms of ontology. The other conclusion is that MaxEnt, as is evident from the direct comparison between the directionality of the pertinent arrows in figs. 2 and 6, is not a physical property of the efficient-causal processes, causing the microstates to move towards macrostate *B*, but is an inference about the microstates that correspond to macrostate *A*. In a sense, this statement just reflects the fact that the meaning of 'disequilibrium' can only be defined relative to an 'equilibrium,' independent from whether the latter could materialize or not.

Now, we can also see the correspondence between the MaxEnt Principle and the notion of selection. The MaxEnt Principle selects a certain macrostate as the one in which all relevant information about an object will have been extracted. The question then becomes how to make the implicit notion of an observer explicit, while maintaining the general notion of selection (again, following Price 1975). Coming back to Short's and R&S's analysis, the notion of purpose can relate to both natural selection and to intentionality, thus enabling us to see a direct correspondence between the notion of a human observer extracting information about an object and a general evolutionary process extracting information about the objects that are involved in that process. This directly corresponds to Maynard Smith's (2000) treatment of the concept of information in biology. Selection is anisotropic, as it favours certain kinds of outcomes -- but not specific ones -- depending on the constraints. Hence the results of selection carry information, but not necessarily with reference to a human observer only. In the MaxEnt framework, selection corresponds to inference as a statistical process. That is -- and corresponding to Maynard Smith's argument -- either the maximum entropy process is related to an observer who intentionally changes her belief about the state of an observed system by utilizing the MaxEnt Principle, or these criteria are generated endogenously by a natural process of selection from one moment to the next. The resulting information is 'intentional' in the very general sense of a reduction of possibilities by way of an historical process (concatenations of causal contingencies) that is independent of reference to a human observer. Clearly -- here we following R&S again -- this requires us to relate the endogenous interpretant to a larger context that we introduced previously as a system of interpretance. That is, fig. 6, when referring to a generalized notion of natural selection as a purely statistical process, would have to be enlarged in the way of fig. 3. Hence, the information generated by a physical interpretant relates to a compositional hierarchy of functions that emerge during the evolutionary process.

This view establishes a direct connection between the notions of evolution and thermodynamics that differs from the commonly held view that evolution works against the Second Law. This common view confuses levels in the sense that the constraints themselves correspond to physical states that are imposing order on the evolving system. For a simple example, material structures emerging from gravitational pull directly manifest the Second Law by adjusting each moment to the most likely next state, but they also represent constraints for other processes, such as chemical ones that might be taking place on their surfaces, as posited in theories about the origin of life. These dependent processes will reflect the constraints, and thus respond to what for them is informative, in spite of the fact that the constraints themselves have emerged spontaneously from a process following the Second Law any which way. Analysis of the autocatalytic cycle is the central paradigm for this, because the cycle as such can be seen as a structure that accumulates information in the sense of restricting its range of actualisations in a space of possible relations. But this view results from a focus on the level of formal causation, and overlooks the factual dynamics on the level of efficient causation, where all the underlying processes tend to maximize entropy production via the maximization of throughput while proximally maximizing power.

As we see from this argument, in the case of the autocatalytic cycle there is a natural extension from the MaxEnt approach to the thermodynamic view in the sense of the physical process of energy dissipation. We will now explore this argument, relying on its generality. Recently, Annila and collaborators (Tuisku et al. 2009; Annila and Kuismanen 2009; Annila and Salthe 2010; Salthe, 2004) have shown that the emergence of hierarchies can be seen to reflect the Second Law because they manifest energy dissipation of available energy gradients in such a way as to tend to maximize entropy production under given physical constraints. This view can be related to figure 7 -- that is, to the notion of a chain of responses / interpretants, which has been postulated by Peirce in the conception of infinite semiosis, especially in his early writings (Atkin 2009). Infinite semiosis follows from the fact that every interpretant can be a sign relative to another interpretant (indicated by the arrows that connect response *R* with signs *X*, relative to the next response *R'*, and so forth). So also, every response is embedded in a hierarchy of responses in living systems -- cellular, organismic and ecological. The increasing complexity of this hierarchy in no wise implies that reference to the ultimate physical purpose – the maximization of entropy production – becomes obviated. All the levels conspire together to maximize local entropy production given the constraints (Salthe 2007). Hence, increasing order, as reflected in the emergence of hierarchies, can be seen to be an expression of the Second Law.

**Figure 7: Infinite semiosis**

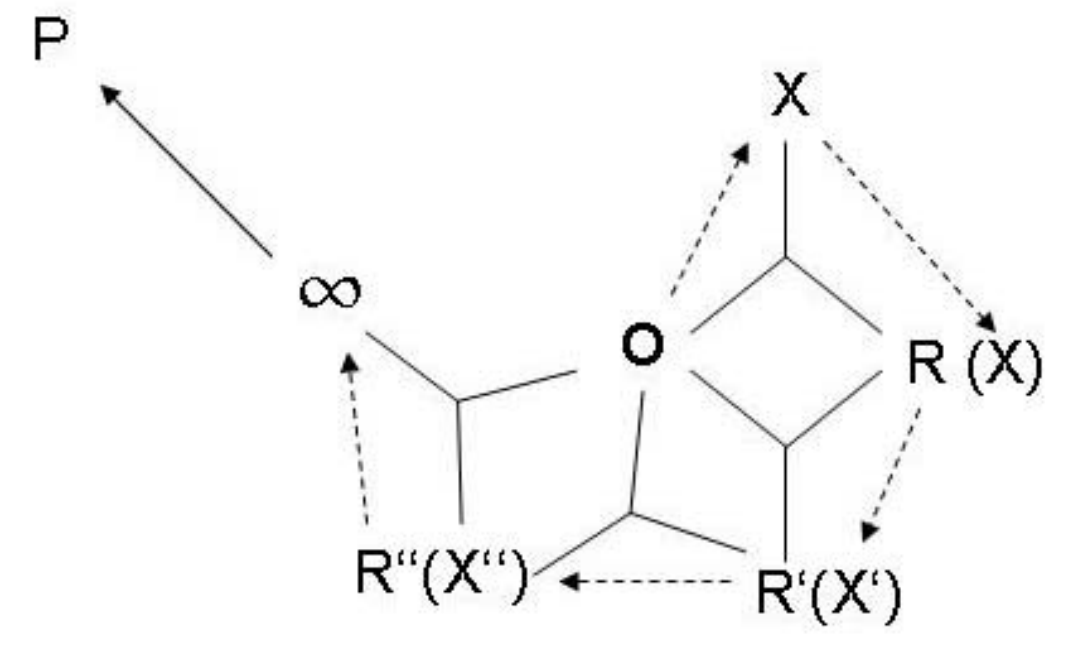

From this follows another Peircean insight. This is that the final macrostate carries information about bearing constraints, but not about the initial macrostate nor about the transient microstates, which are just drawn sequentially from among various most probable ones locally. The information about macrostate *A* is lost because the system could arrive at macrostate *B* from any number of possible macrostates caused by microstates far from the maximum entropy configuration, and the information about the microstates is lost because the microstates have maximized their informational entropy from any prior configurations. This provides a deeper insight into why Peirce distinguishes between the final and the dynamical interpretants, and, in his later writings no longer focused on infinite semiosis, being mainly concerned with the question how ultimately the reality of the object is constituted in the semiotic process, which corresponds to the question how information emerges from that process. However, if a configuration carries information, it could be interpreted as a sign by another process, which in turn accumulates information about the constraints under which that process proceeds. So, a claim that a triadic process generates information is inexact because a single triadic thermodynamic process would not generate any information at all, but only maximizes informational entropy (informational carrying capacity). The information embodied in the final state can only be actualised in relation to another triadic process, in which the interpretant (qua equilibrium state of the former process) functions as a sign. This corresponds to our previous analysis that it is not the smoke that embodies the information about the fire, but the responses of other systems to the smoke, all embedded in a system of interpretance. In this fundamental sense, the semiotic relation between thermodynamics and evolution can only be discerned if the process of evolution is considered in its entirety, and not just during an arbitrarily selected single step.

### *3.3. The thermodynamics of semiosis*

We can now invert the direction of the analysis and apply the thermodynamic perspective to semiosis. This was already implicit in interpreting the macrostate at thermodynamic equilibrium as a final interpretant. Following our analysis of the autocatalytic cycle, this leads us to ask three questions.

- First, can we interpret the relation between the response, R, and the object, O, in terms of the Maximum Entropy principle?
- Second, does the application of the MaxEnt principle here have a physical meaning?
- Third, if so, what is the relation between MaxEnt and the Maximum Entropy Production Principle (MEPP)?

Regarding the first, we can directly rely on Dewar's (2005, 2009) recent explication of MaxEnt as an inference process that corresponds to Jaynes' classical treatment of entropy. Dewar uses this argument to show that MaxEnt is a methodological approach by which a physical theory can identify the constraints that operate on an observed physical system, such that the information loss from all unobserved properties is minimized, validating the 'other things being equal' stipulation. The MaxEnt state, given the constraints, means that it is unnecessary to have more information about those states than what the equiprobability assumption implies -- i.e. Laplace's 'principle of indifference'. This argument can be naturalized, if we consider that the observer can be any sort of evolving physical system, which we might call an 'inference device' (Wolpert 2008) – an entity that stands in a general relation of observation with what is therefore an observed system.

**Fig. 8: MaxEnt and selection**

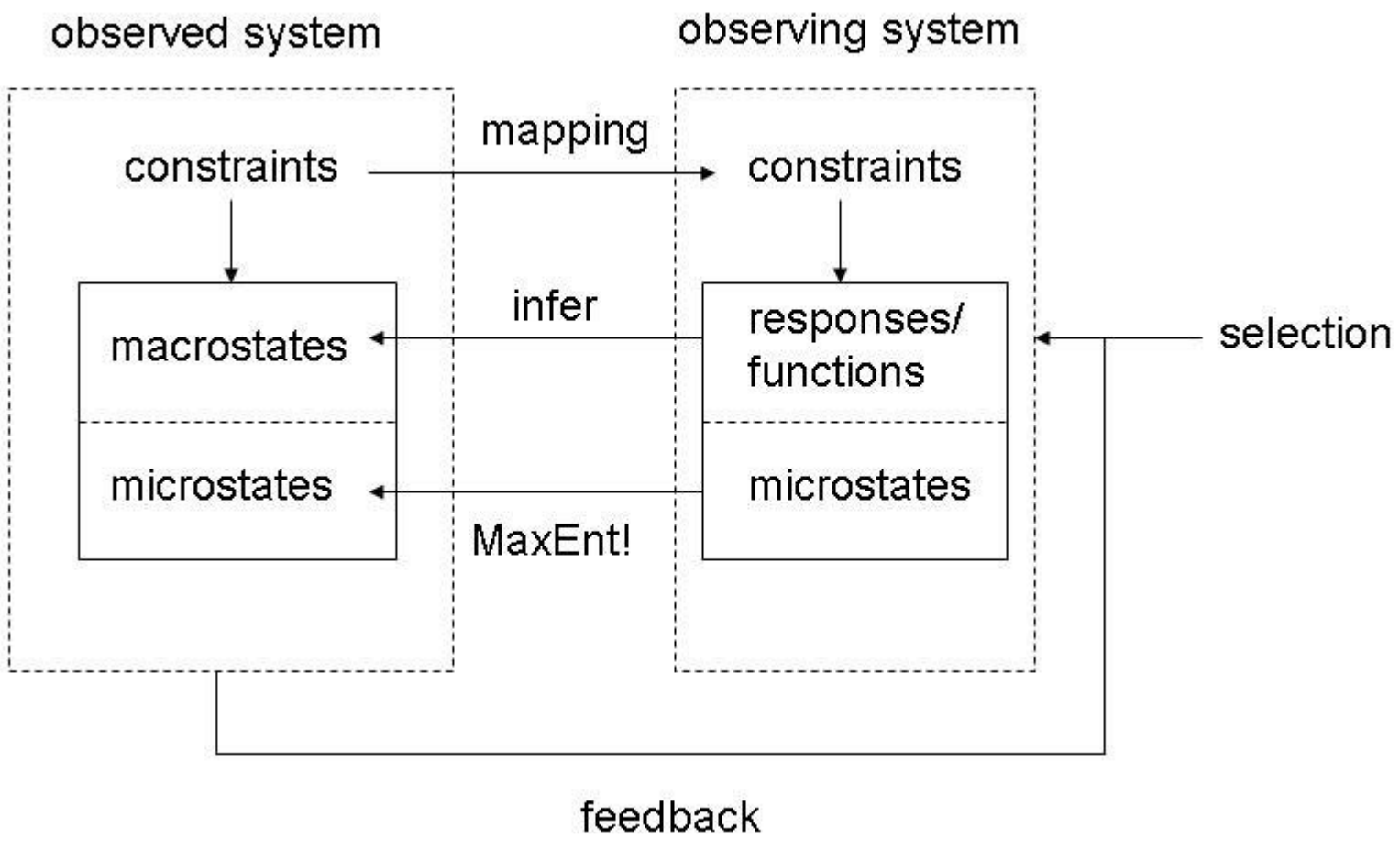

This inference device is a system of functions that trigger responses to observations (see fig. 8). In the evolutionary turn of the argument, it is important that the observing system is a member of a population of systems competing under natural selection. In this perspective, we can interpret the Dewar argument as describing the results of selection that has acted upon a population of observing systems. Any sort of physical inference device, as a member of a selected population of them, will tend towards a state in which the constraints operating on the observed system will be inferred from a particular macrostate configuration such that the inference device will in all other respects manifest a maximum entropy state with regard to all possible mappings between microstates of the observed system and microstates of the inference device. In this way the macrostate configuration of the object acts as the sign in a triadic relationship, which is mapped onto macrostates of the observing system, i.e. the inference device. Thus, constraints on the observed system are turned into constraints operating on the observing system, defining certain evolving structural properties, which in turn function as objects for other observing systems, thus constituting evolution as a sequence of semioses.

**Fig. 9: The semiotic view on MaxEnt**

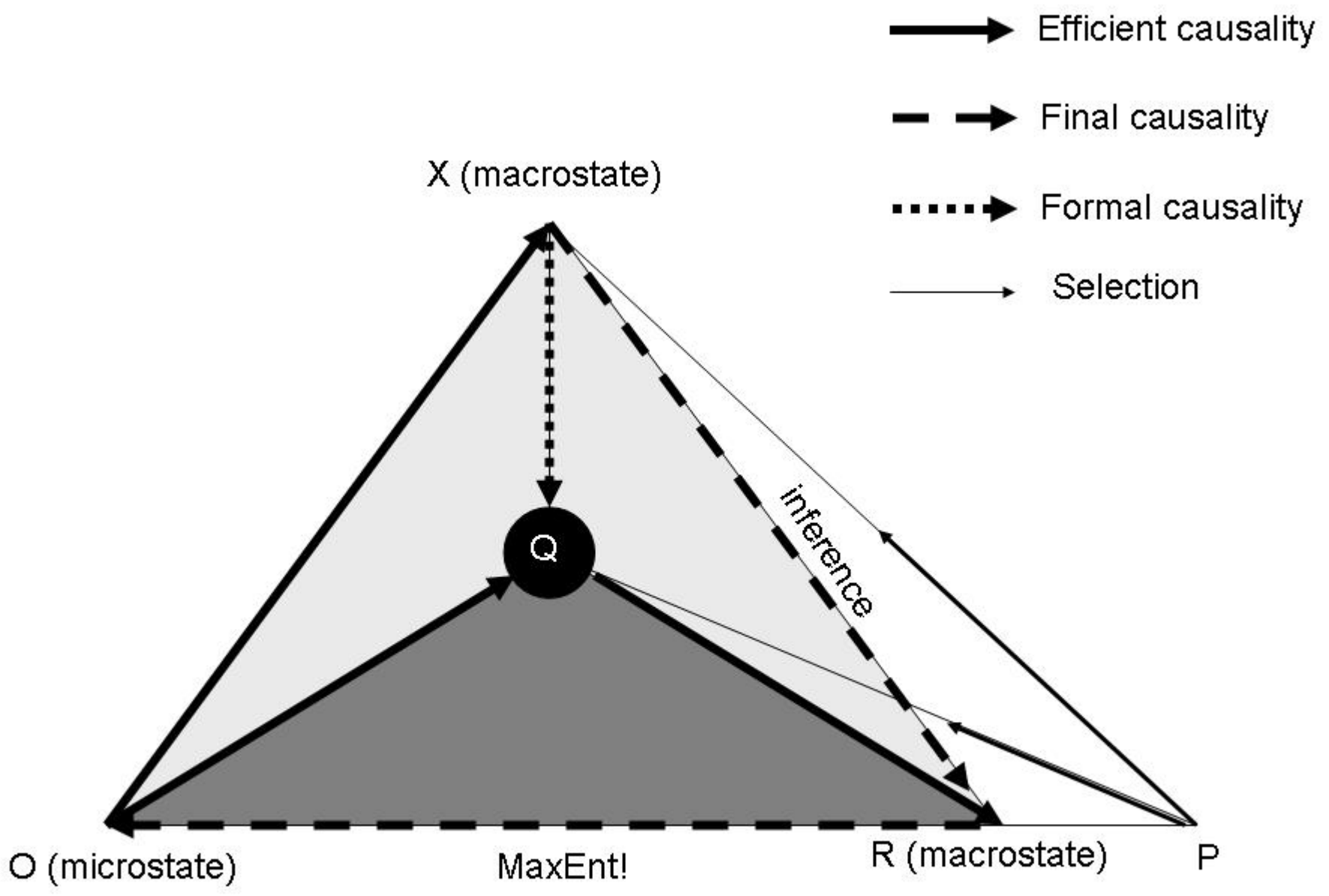

Now we can project this view onto the semiotic triad (fig. 9). The labels are as in figures 2 and 4. The observing / inferring system under selection (*Q*, *R*) constructs a macrostate as a sign (*X*), which guides its inferences and responses (interpretants) with respect to the object (*O*). In other respects the system assumes a maximum entropy stance of readiness to react to changes in its relation with O. Then we can see that the MaxEnt hypothesis is an hypothesis about the co-evolution of the object and *Q*, which changes its structure under natural selection. In a population of entities with the property *Q*, this property will show shared constraints, reflecting constraints operating upon the object system, and a maximum variety in all other respects. In the context of biology, this is the relation between species' characteristics, related to the relevant ecological niche, and the irreducible individuality found among the members of the population of a species. Therefore, we can conclude that the biological notion of a 'species' is a semiotic category, and corresponds to the role of the autocatalytic structure in our analysis above. And so the species, as does an autocatalytic cycle, functions as a formal cause. Thus, the semiotic analysis allows for a reinterpretation of the perennial tension between the adaptationist - population genetic concept of a species (which reflects the strand of efficient causality in the semiotic model) and the taxonomic - morphological ones (which reflect the formal and final causality in the semiotic model).

This insight is not new, viewed from another angle, because it is another expression of Ashby's (1958) principle of 'requisite variety'. The inference device will evolve into a structure that reflects external constraints on object systems, and at the same time will maximize information capacity, hence the possibility to generate variety, which also maximizes its future potential for information generation – to evolve. This was the major thrust of the Dobzhansky school of the 'modern synthesis' (Lewontin, 1974). So, we can relate the MaxEnt principle with fundamental assumptions about the evolutionary process, in particular with the idea that evolution simultaneously drives the emergence of order and permits an exuberance of the variety of life forms. Recently, Frank (2009a,b) has provided the analytical foundations for this conception in showing that different statistical distributions in nature can be explained as expressions of the maximum entropy approach in nature. This reasoning supposes that evolution is a statistical process which interacts with other phenomena in nature that also are stochastic, such as variations in climate. The general idea is that, as a result, statistical distributions of biological phenomena reflect the constraints that are imposed on living systems by environmental conditions, such that all fluctuations tend toward the maximum entropy state. Thus, the statistical properties of living systems, such as the distribution of species in an ecosystem, can be explained by the MaxEnt principle (for related approaches, see Dewar and Porté 2008, and Grönholm and Annila 2007).

Another approach that is apposite to ours is Ulanowicz's (Ulanowicz 1997, Ulanowicz et al. 2009) information-theoretic conception of ecological 'ascendency.' Ascendency relates the mutual information of matter-energy flows in an ecosystem -- hence a measure of the constraints (which correspond to *Q* in our conceptual model) exerted upon those flows -- to the total system throughput. Ascendency, numerically determined as the product of TST (Total System Throughput) and AMI (Average Mutual Information), is one part of the total developmental capacity of a system, which also depends on the available information capacity, that is, the 'unorganised complexity' of a system. Clearly, this corresponds to the notion of constraints versus the maximum entropy postulate in our semiotic approach. Ulanowicz (Ulanowicz and Hannon, 1987), emphasize the role of living systems in creating a tendency to maximal energy dissipation across all hierarchical levels, even though this is not necessarily the case for any given system under particular environmental conditions, such as strongly resource-constrained ones (Meysman and Bruers 2010). Hence, it is central to observe the conceptual consistency and coherence of the maximum entropy approach, which implies that the principle itself has to be conceived as a probabilistic one, which necessarily implies that there will be spatial and temporal deviations from it. Yet, this also means that, for example, maximum entropy production will hold over the longer time spans that cover the evolutionary trajectories of living systems (Vallino 2010), or, that there is an increasing variability of the distribution of flux gradients, and that entropy production is maximized over this entire range (Niven 2010). Indeed, the standard argument that complexity counters the sway of the Second Law can be summarily disposed of in considering the entire tree of life reflecting one stochastic fluctuation under certain constraints (Gould 2003: 899ff.), such that this fluctuation is the relevant unit of thermodynamic analysis.

In our thermodynamic view on the semiotic triad, one piece is still missing. This involves the proper interpretation of the $P$ in figure 9 – that is, of the role of selection. This leads us to the contested issue whether MaxEnt also implies the Maximum Entropy Production Principle. It is central to see that we have already taken MaxEnt beyond being a mere methodological principle, because we assume that MaxEnt entails the maximization of the informational entropy of physical states of observing systems. In Salthe's (1993) terms, this corresponds to an internal entropy, or observer relative entropy (Herrmann-Pillath 2010). Obviously, the observing system's entropy is not the same as the entropy of the entire ensemble of observing and observed system, or, as the physical entropy production of the complete semiotic process, a distinction which is easy to overlook and which lies at the root of the long-standing controversies about Maxwell's demon (Maroney 2009).

The answer to the question of how MaxEnt relates with MEPP requires analysis of the causal relationship between the observed system and the observing system. This is the reverse side of the inference process. Inferences result from causal interactions, as in the case of a bacterium that searches for nutrients and follows certain chemical gradients which emanate from the nutrients.

A straightforward way to theorize this causal relationship is to use the maximum power principle. The MPP has been made a cornerstone of evolutionary analysis by Vermeij (2004) recently, and directly follows a long sequence of contributions going back to Lotka's original contribution (1922a,b), such as Odum (2008). It states that living systems will tend to maximize power (energy throughput) during their causal interactions with the environment, while also being subject to natural selection. Salthe (1975) presented evidence that selection is most intense during intense behavioral activity. A related argument has been deployed in positing the so-called 'constructal law,' which states that physical flow systems will always evolve into states in which access to the energy gradients is maximized (Bejan and Lorente 2006, 2010). MEP occurs over the entire time scale of the processes of a flow system, with the intermediate stage of maximum power production, at which a tendency of minimization of dissipation holds place, thus driving the efficacy of work generated during the life cycle of the flow system. That means, MPP can be seen as a roundabout way in the manifestation of MEPP, which ultimately realizes states with higher entropy production, compared to states that do not manifest the structural properties underlying maximization of flows.

We can establish a conceptual relation between MPP and MaxEnt by means of the distinction between physical work and dissipation. MPP refers to the maximization of exergy consumption in a useful way. That is, we can distinguish between two levels of the causal interaction between two systems. One is the macroscopic interaction, which we assimilate to physical work, and the other is the microscopic interaction, which we define as dissipation. All causal interactions are energy transfers (Bunge 1977: 240, 326). Those transfers can have two shapes. One associates with transfers that relate a macroscopic property of one system with a macroscopic property of another system; this is classified as 'work.' The other is the microscopic interaction -- dissipation, such as via the friction co-occurring with the macroscopic interaction. The central point is that the categorization of a causal impact as 'work' is relative to the macrostates of the receiving system in the causal interaction.

Now, the point about MPP is that, under natural selection, a system will evolve in the direction of maximizing the energy utilization for work in its own interactions with the environment, that is, maximizing the energy flows that carry the exergy used to produce useful work, with the criteria of 'usefulness' having been determined by the natural selection of functions.

This statement relates to MaxEnt because it implies a direct connection with the inference process. Maximizing power requires a tendency to approach a global macroscopic state in which the relevant macroscopic states of the object system are identified, and such that all other causal interactions with the object system become temporarily irrelevant -- that is, they tend towards the maximum entropy state. In other words, under natural selection, a system will tend towards maximizing its derivation of exergy from other systems while minimizing dissipation. Now, as is well known from related theorems in engineering, this relation results in an optimum energy flow rate (not the most efficient rate) (for applications of this principle in the earth sciences, see Kleidon 2009, 2010). Together with the ongoing dissipative processes, we conclude that the entire process chain results in maximum entropy production to the extent possible given the constraints. That is, the system has become organized by selection in such a way that increased power can be achieved by increasing work rate up to a point, and increased power necessarily associated as well with increased dissipation over the entire developmental trajectory of the system exerting work on its environment. Maximum power is therefore the avatar of Maximum Entropy Production in complex systems.

We summarize these connections in figure 10. This shows that the centerpiece of an MEPP process is the semiotic process of the observing system. This results from three determinants. One is that MEPP can be regarded as the ultimate purpose that reigns over the natural selection of any local purposes (as in figure 9), as well as over the various intermediate functions in systems of interpretance connected via infinite semiosis. This materializes in two different processes. One is MPP, which directly follows from the workings of natural selection, following Lotka's conjecture, and related theories, such as Bejan's constructal law approach. The other is MaxEnt, which here functions as the principle of preserving and increasing information capacity during natural selection. No adaptation can be complete and perfect, but leaves a system with overhead for flexible responses, shown in figure 10 as the 'accumulation of information capacity', a point made often by Lewontin in regard to genetic variability in populations (e.g. 2003).

**Figure 10: Relationship between MaxEnt, MPP and MEPP**

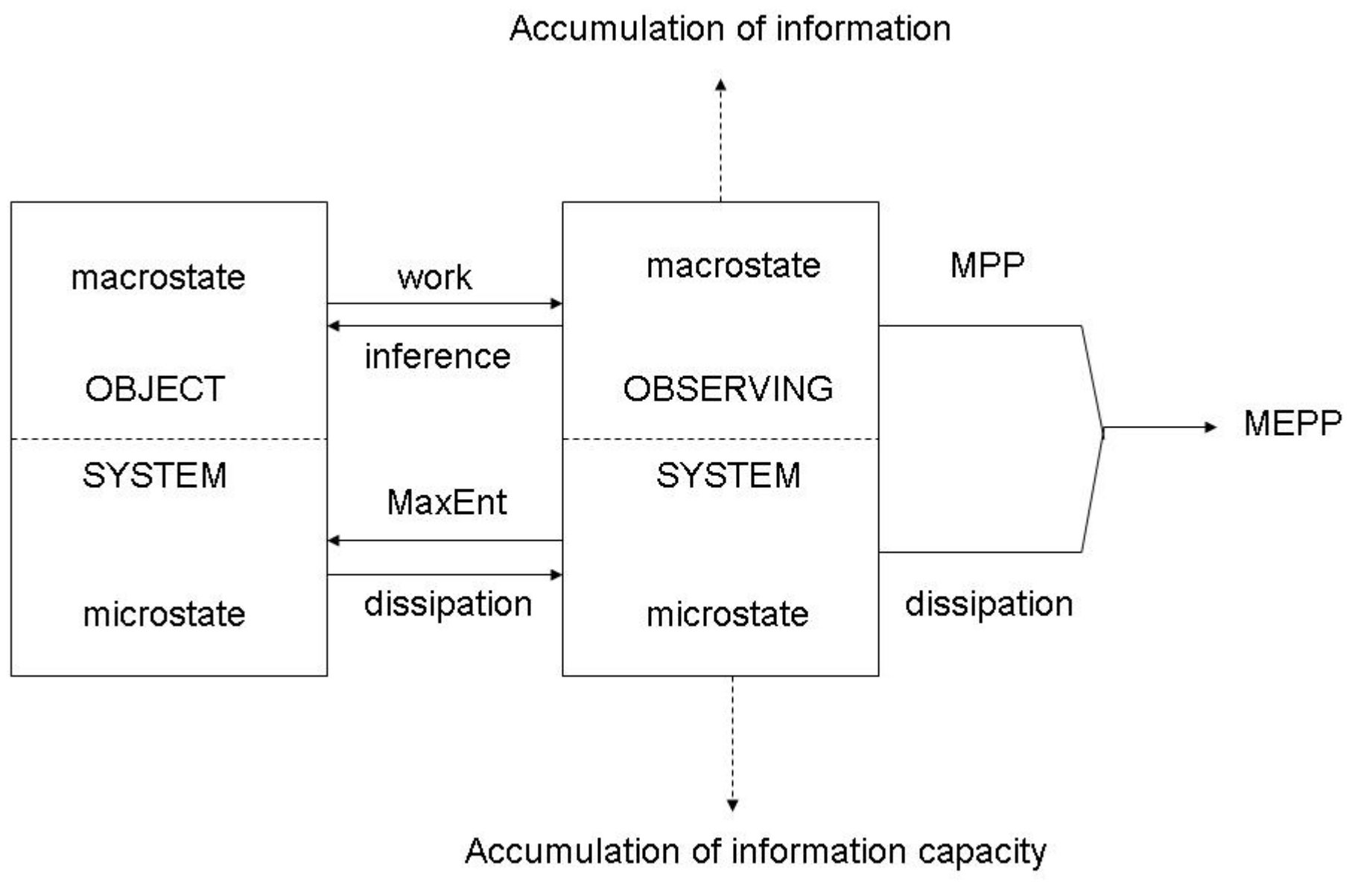

From this overview we conclude that the standard view, originating from Schrödinger (1944), that life is a phenomenon counteracting the Second Law, is wrong, because it focuses only on the macroscopic side and the aspect of information generation. The entire process in fact follows the Second Law, because the emerging semiotic structures speed up energy dissipation during the marshalling of ever more complex intermediating structures during semiosis. This process can be compared to that of capital formation in the economy, as well as with the functions of human technology. All of these can, in the context of a disequilibrated universe, be viewed most generally as occasions for bringing the universe closer to thermodynamic equilibrium in situations where energy gradients do not spontaneously dissipate by conduction. Clearly, compared with the agricultural economies of the past, modern industrial technologies harness vastly larger amounts of exergy, much of it having been sequestered deep in the earth. This is made possible via capital formation, that is, investing in more complex intermediating structures, which also reflect the rapid accumulation of knowledge about the management of physical processes. So, in the end human technological systems maximize energy throughput, which, in spite of all improvements in energetic efficiency, and economic growth reflects the working of the Second Law (Ayres and Warr 2003, 2005; Annila and Salthe 2009).

This approach allows us to clarify a discussion in the literature in which the process of evolution is seen as generating information that harnesses energy for the creation and maintenance of ordered structures -- for example, Corning's (2005) notion of 'control information.' Our point here is that, while control information may require much less energy throughput to be maintained than the energy throughput that is actualised by means of it, this does not contradict the argument presented here, because it implies that MPP holds for the system of which embodied control information is a part. In other words, it is MPP that is actualised by means of control information, and this pushes MEPP to the limits possible in these structures, leaving unrealized energy flows and systemic collapse (and lost MEPP) as the default option.

So, we can summarize the relation between the different principles in fig. 11. This reveals the fundamental division between the internal and external concepts of the actualisation of entropy increase. These distinctions are indispensable for understanding the role of semiosis under the influence of the Second Law, and which at the same time entails an important epistemological conclusion. This is that the scientific observer, as part of the interactive systems, cannot directly access the information needed to measure the flow of entropy generated according to the Second Law. A living system can only operate below the line separating internal and external concepts of entropy in the figure. This conclusion corresponds to the Jaynes conception of entropy, which states, in a famous expression, that "entropy is an anthropomorphic concept" (Jaynes 1965). Jaynes, in his framework, refers to the fact that there is no unified measure of entropy across all possible experimental configurations, which impose different kinds of constraints upon physical systems. This is a general feature also of information-theoretic measures of informational entropy, which always explicitly refer to an observer and the corresponding partitioning of the state space (Kåhre 2002: 181ff.).

**Fig. 11: The Second Law in its relation to MEPP, MPP and MaxEnt**

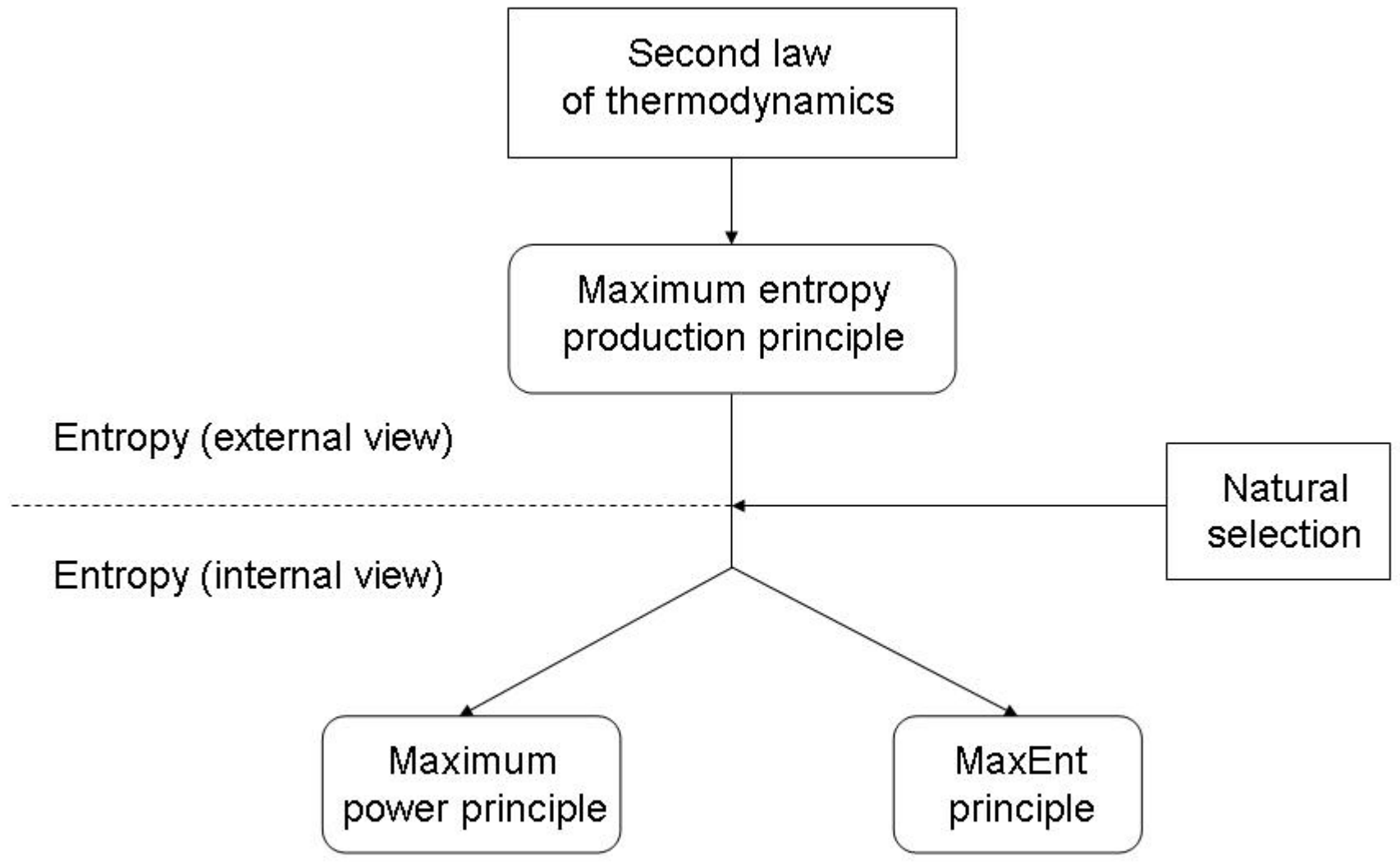

So, this conjunction of the different entropy related theorems can serve as the theoretical framework in which the general physical regularities of information-generating evolution can be analyzed, extending and detailing earlier cosmological approaches such as that of Chaisson (2001). In this framework, life is not seen as counteracting the Second Law, but as expressing it through the emergence of structures which increase the speed and efficacy of entropy production. This view concurs with non-equilibrium thermodynamics, but adds the perspective of semiotics. Only in the semiotics view can we understand how the dissipation of energy relates with the accumulation of information. We think that a philosophical approach following Peirce will be indispensable in order to complete the recently emerging 'physics of information' paradigm.

## 4. Conclusion

The transition from dyadic to triadic conceptual structures is a fundamental methodological shift that allows for the cogent analysis of the conjunction of different forms of causality in information-generating evolution. In his original contributions, Peirce consistently emphasized the distinction between ‘Firstness,’ ‘Secondness’ and ‘Thirdness,’ which for some may seem to be a conceptual panacea to accommodate almost any thinkable theoretical approach, given the open references of the terms that are exploited by Peirce himself (Burch 2010). In this paper, we elect to relate the triadic framework to the conjunction of efficient, formal and final causality in the semiotic process. This is possible if we follow the path opened up by Peirce himself, namely understanding semiosis as a physical phenomenon. This is possible if we eschew the Cartesian reference of ‘interpretation’ to mental substances, and view the Peircean interpretant as ‘responses’ or functions of evolving systems.

We think that triadic conceptual structures will help to clarify many open issues in evolutionary theory, especially in terms of reintegrating alternative approaches that emerged in the past decades, which continue to struggle with the widespread perception that evolutionary theory is about another kind of causality (Hull et al. 2001), a position that is as old as Darwinian theory itself (Blitz 1992). For example, in the Peircean framework it is straightforward to relate natural selection in the narrow NeoDarwinian sense (which conceptually builds on efficient causality alone) to the theory of signal selection proposed by the Zahavis (Zahavi and Zahavi 1997), which even a staunch NeoDarwinian like Dawkins (1989) had perceived as a major paradigmatic challenge. Signal selection builds on a synthesis of meaning and function in the analysis of evolving traits in living systems, and thus implicitly introduces a triadic viewpoint, fusing two alternative dualisms in the analysis of adaptation, namely organism / environment and organism / organism. Similarly, the theory of niche construction (Odling-Smee et al. 2003) acquires a straightforward interpretation in the light of triadic structures, as it merges two other dualisms, namely adaptation of organism to environment and adaptation of environment to organism. As a final example, in recent extensions of the Darwinian paradigm, the ‘interpretive’ functions of the cell and of the organismic environment in mediating gene expression have been demonstrated empirically, with the possible consequence of merging the theories of biological and cultural evolution (Jablonka and Lamb 2006). Putting different evolutionary processes into the universal framework of Peircean semiotics should help to clarify important issues in these debates.

However, as we have seen, the triadic framework also changes even more foundational assumptions about the physics of evolution. This is because it offers a new view on the relation between statistical and phenomenological thermodynamics. We have generalized over the Bayesian approach to statistical mechanics in the sense of resolving what could be called a ‘second-order mind projection fallacy’. Jaynes (2003) had argued that assigning the status of objective propensities to probabilities suffers from that fallacy, that is, projecting states of knowledge of the observer onto the real physical world. We radicalize this idea by substituting for the 'mind', i.e. the observer, an information-generating evolutionary process. This turns 'mind' into a physical phenomenon that is part and parcel of the physical systems under scrutiny. Specifically, this step allows for a conceptual integration of the MaxEnt principle, which is Bayesian in origin, with the physical MEPP.

One of the fundamental problems tackled by Peirce was the question of how regularities become possible in a random world. Use of his triadic structure shows that this "habit formation" happens when a random fluctuation relates causally with a process that is entrained by final causation. This is the case if the environment in which a random fluctuation takes place imposes constraints on (or contextualizes) further change, which then becomes manifest in the consequences of that fluctuation, which maximizes entropy production in the process of reflecting those constraints. This view comes close to Bateson's (1971) classical definition of information as a difference that makes a difference (to some system of interpretance). A random fluctuation would have different consequences against different background conditions, since the fluctuation causes responses that could go off in different directions according to conditions. Thus, the difference becomes manifest in some system of a higher logical type than the random variation, and this corresponds to the notions of both 'macrostate' and 'sign.' This perspective has also been taken in recent attempts to interpret basic quantum processes and information processes, which see physical evolution as a result of quantum computation (Lloyd 2006). Peirce himself might be viewed as having implicitly followed such a model when trying to understand how 'habits' emerge from randomness.

We can summarize this conclusion in a 'universal triad' (fig. 12) which shows the relation between the three kinds of causality, as discussed in the course of our argument.

**Figure 12: The universal triad**

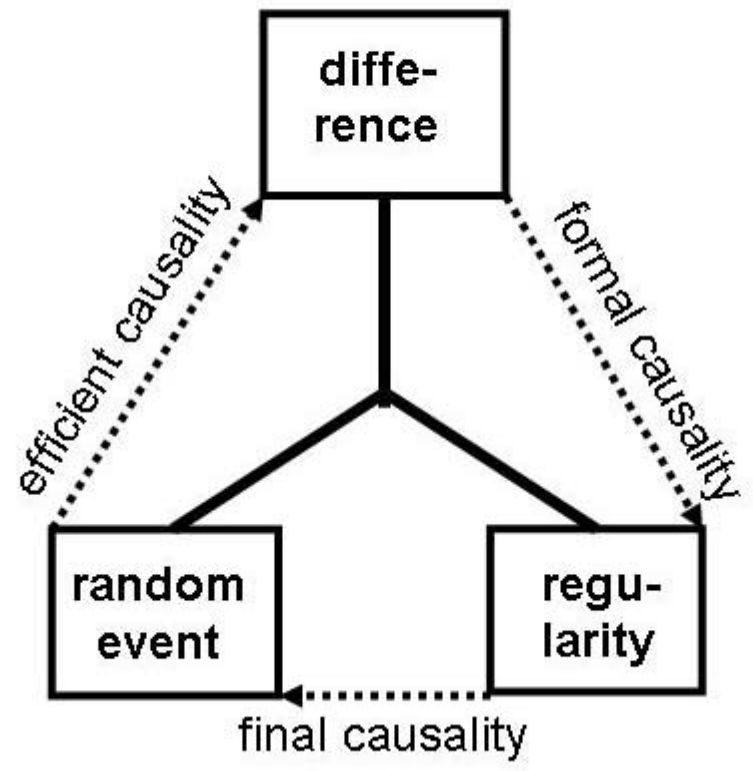